\documentclass[prr,aps,superscriptaddress,twocolumn,10pt]{revtex4-2}

\usepackage{tikz}
\usetikzlibrary{%
    decorations.pathreplacing,%
    decorations.pathmorphing,%
    arrows
}
\usetikzlibrary{positioning, calc,intersections,shapes,matrix,fit}

\usepackage{float}
\usepackage{graphicx}
\usepackage{dcolumn}
\usepackage{bm}
\usepackage{xcolor}
\usepackage{appendix}
\usepackage{caption}
\usepackage{subcaption}
\usepackage{multirow}
\usepackage{booktabs}
\usepackage{amsmath}
\usepackage{amsfonts}
\usepackage[braket, qm]{qcircuit}
\usepackage[normalem]{ulem}
\usepackage{url}
\usepackage{array}
\usepackage{titlesec}
\usepackage{etoolbox}

\usepackage{amsthm}
\newtheorem{theorem}{Theorem}
\newtheorem{lemma}[theorem]{Lemma}

\theoremstyle{definition}
\usepackage[english]{babel}

\newcommand{\afchem}{Department of Chemistry, University of California, Berkeley, California 94720, USA}
\newcommand{\afbqic}{Berkeley Center for Quantum Information and Computation, Berkeley, California 94720, USA}

\begin{document}
\preprint{APS/123-QED}

\title{An Error Mitigated Non-Orthogonal Quantum Eigensolver via Shadow Tomography}

\author{Hang Ren}
\affiliation{\afbqic}
\affiliation{\afchem}
\author{Yipei Zhang}
\affiliation{\afbqic}
\affiliation{\afchem}
\author{Wendy M. Billings}
\affiliation{\afbqic}
\affiliation{\afchem}
\author{Rebecca Tomann}
\affiliation{\afbqic}
\affiliation{\afchem}
\author{Nikolay V. Tkachenko}
\affiliation{\afbqic}
\affiliation{\afchem}
\author{Martin Head-Gordon}
\affiliation{\afchem}
\author{K. Birgitta Whaley}
\email{whaley@berkeley.edu}
\affiliation{\afbqic}
\affiliation{\afchem}
\date{\today}

\begin{abstract}
We present a shadow-tomography-enhanced Non-Orthogonal Quantum Eigensolver (NOQE) for more efficient and accurate electronic structure calculations on near-term quantum devices. By integrating shadow tomography into the NOQE, the measurement cost scales linearly rather than quadratically with the number of reference states, while also reducing the required qubits and circuit depth by half. This approach enables extraction of all matrix elements via randomized measurements and classical postprocessing. We analyze its sample complexity and show that, for small systems, it remains constant in the high-precision regime, while for larger systems, it scales linearly with the system size.
We further apply shadow-based error mitigation to suppress noise-induced bias without increasing quantum resources. Demonstrations on the hydrogen molecule in the strongly correlated regime achieve chemical accuracy under realistic noise, showing that our method is both resource-efficient and noise-resilient for practical quantum chemistry simulations in the near term.
\end{abstract}

\maketitle

\section{Introduction}
Quantum computing offers the potential for simulating molecular electronic structures that are intractable with classical methods \cite{mcardle2020quantum, jordan1993paulische}. Accurately determining the ground and excited state energies of molecules is central to understanding chemical reactions, catalysis, and materials design. While variational quantum eigensolvers (VQEs) have received significant interest for these problems, the iterative optimization that lies at the core of these methods often becomes measurement and resource-intensive as system size grows \cite{cerezo2021variational, tilly2022variational}, and may be constrained by the phenomenon of barren plateaux in high dimensions~\cite{larocca2024review}. Non-variational methods such as Quantum Phase Estimation (QPE) \cite{kitaev1995quantum, abrams1997simulation} typically require significantly deeper circuits that exceed the capabilities of current quantum hardware. 

The Non-Orthogonal Quantum Eigensolver (NOQE) \cite{baek2023say} 
presents an alternative approach that has been shown to provide significant quantum advantage for electronic systems with both strong and weak electron correlations. By employing multiple non-orthogonal reference states derived from unrestricted Hartree-Fock (UHF) solutions \cite{pratt1956unrestricted} and a unitary coupled cluster (UCC) ansatz, NOQE solves a generalized eigenvalue problem \cite{ghojogh2019eigenvalue} to obtain molecular energies directly, thereby circumventing costly variational optimization. The use of multiple reference states allows description of strong correlation, while dressing by UCC-type ansatze allows incorporation of week correlations. NOQE is both variational and size consistent, two key desirable features found in classical electronic structure algorithms only for the uncorrelated Hartree-Fock solution, and for full configuration interaction, which is intractable for large systems. The classical analog of NOQE, namely non-orthogonal configuration interaction (NOCI) with exponential correlators, is intractable and the closest classical analogs are exponentially more costly than NOQE~\cite{baek2023say}.

While NOQE avoids the high cost of repeated measurements associated with variational optimization, measuring the matrix elements can still be expensive, given that matrix size scales quadratically $\mathcal{O} (M^2)$ with the number of non-orthogonal reference states $M$. Additionally, the estimation of individual Hamiltonian matrix elements requires measuring a series of Pauli strings, which may be large for systems with complex Hamiltonians. 

A number of protocols have been proposed to simplify measurement strategies of electronic Hamiltonians \cite{o2019generalized, verteletskyi2020measurement, rubin2018application, huggins2021efficient, van2024end}. These have typically focused on diagonal elements of the Hamiltonian matrix, because of the emphasis in the community on VQE methodologies. However in NOQE we require both diagonal and off-diagonal Hamiltonian elements, as well as off-diagonal overlap elements between non-orthogonal reference states. For diagonal Hamiltonian matrix elements, the most efficient protocols require additional unitary operations to generate pre-measurement entanglement \cite{huggins2021efficient} and are not directly applicable to measurement of off-diagonal matrix elements.

In this work, we address these issues by integrating shadow tomography (classical shadows for quantum states) \cite{huang2020predicting} into the NOQE framework. Shadow tomography is a powerful measurement protocol that efficiently predicts multiple properties of a quantum state from a small number of randomized measurements. By applying shadow tomography to each reference state and its phase-shifted superpositions with the vacuum state (auxiliary variants), we eliminate the need to jointly prepare pairs of reference states. Instead, the necessary Hamiltonian and overlap matrix elements are reconstructed classically from classical shadows. This reduces the measurement scaling with respect to the reference states to $O(M)$. It also halves the number of required qubits $N$ and shortens the circuit depth, making the protocol more experimentally feasible.

An additional advantage of our approach is its natural compatibility with shadow-based error mitigation techniques, such as shadow distillation \cite{PRXQuantum.4.010303}. This enables the suppression of noise-induced bias in the estimated energies without incurring additional quantum overhead.

We first analytically demonstrate the effectiveness of the shadow-tomography-enhanced NOQE method by analyzing its sampling complexity and circuit size, and comparing them to those of the standard NOQE protocol using Hadamard measurements. We then validate its practical performance through an application to computing the electronic energies of the hydrogen molecule near the Coulson–Fischer point, a regime characterized by strong electron correlation. In this application, we use noise models based on the state-of-the-art Quantinuum Model H2 hardware system~\cite{quantinuum_h2_2024}. 

These demonstrations show that shadow-tomography-enhanced NOQE provides a practical and noise-resilient framework for realizing accurate electronic structure calculations, making it suitable for implementation on near-term quantum devices.

The remainder of the paper is organized as follows. Section \ref{NOQE} reviews the NOQE framework. Section \ref{shadow} introduces the fundamentals of shadow tomography. Building on these concepts, Section \ref{sec:Applying Shadow Tomography to NOQE} details how to integrate shadow tomography into NOQE, and Section \ref{sec: resource reduction analysis} undertakes a sampling complexity analysis that quantifies the resource requirements for measurements using shadow tomography and for using the Hadamard test approach of Ref.~\cite{baek2023say}. In Section \ref{shadow distillation}, we discuss how shadow distillation can be implemented as an error mitigation strategy for NOQE. Section \ref{evaluation} demonstrates the accuracy and efficiency of our approach with calculations for electronic states and energies of the hydrogen molecule, while Section \ref{sec:Performance of the protocol with noise} confirms the effectiveness of the protocols under realistic noise. Finally, Section \ref{sec:discussion} summarizes our findings and outlines potential directions for future research.

\section{NOQE Framework analysis}\label{NOQE}

This section reviews the Non-Orthogonal Quantum Eigensolver (NOQE) framework and its measurement requirements. In Sec.\ref{sec: Working principle of NOQE}, we outline the underlying principles of NOQE. In Sec.\ref{sec: original}, we summarize the original NOQE circuit that measures Hamiltonian and overlap matrix elements. Finally, in Sec.~\ref{sec: sample complexity}, we quantify the sample complexity of the original protocol, providing a baseline for comparing our shadow tomography-based approach.

\subsection{Working principle of NOQE}\label{sec: Working principle of NOQE}

Accurate electronic energy estimation is central to quantum chemistry. Traditional algorithms such as the VQE can approximate ground-state energies but often require extensive iterative optimization and very large numbers of measurements \cite{huggins2020non}. The NOQE offers a different approach, eliminating the need for 
expensive variational optimization by using a set of non-orthogonal reference states to project the Hamiltonian into a smaller subspace and then directly solve a generalized eigenvalue problem in that subspace.

To construct these reference states, NOQE applies a unitary coupled-cluster (UCC) correlator $e^{\hat{\tau}_i}$, e.g., the UCCD correlator that includes double excitatons, to an unrestricted Hartree-Fock (UHF) state $|\psi_{\mathrm{HF}}\rangle$: \begin{equation}\label{eq: ucc+hf} |\psi_i\rangle = e^{\hat{\tau}_i}|\psi_{\mathrm{HF}}\rangle. \end{equation}
(see Appendix \ref{sec:Decomposition of Ansatzes} for details).  A key feature of the NOQE approach is that these reference states deriving from distinct UHF states are not orthogonal.

The UHF is chosen over restricted Hartree-Fock (RHF) for its ability to introduce symmetry-broken references and thus achieve overall greater flexibility in description of strongly correlated systems. The UHF reference offers a mean-field approximation to the wavefunction of the system and weak (dynamic) correlation is introduced by the UCCD dressing. The strong (static) correlation is captured through the inclusion of multiple symmetry-broken reference states.

The variational parameters of the UCCD ansatz can be approximated through classical preprocessing
at polynomial cost using second-order many-body perturbation theory \cite{baek2023say}. Relatively shallow circuit implementation of the UCCD ansatz can be realized through a low-rank tensor decomposition of the cluster operator and consequent trotterization \cite{motta2021low}. We focus here on the UCCD ansatz but other alternatives such as the unitary cluster-Jastrow ansatz that can reduce circuit complexity are also possible~\cite{baek2023say}.


The NOQE eigenstates are expressed as a linear combination of dressed non-orthogonal reference states: \begin{equation} |\Phi_{\mathrm{NOQE}}^{k}\rangle = \sum_{i=1}^M c_i^k |\psi_i\rangle. \end{equation} 
The coefficients ${c_i^k}$ are found by solving the generalized eigenvalue problem \begin{equation}\label{gep} \mathbf{H}\vec{c}^k = E^k \mathbf{S}\vec{c}^k, \end{equation} where matrix elements of the Hamiltonian and overlap matrices $\mathbf{H}$ and $\mathbf{S}$ are given by
\begin{equation}\label{matrices} H_{ij} = \langle \psi_i|\hat{H}|\psi_j\rangle, \quad S_{ij} = \langle \psi_i|\psi_j\rangle. \end{equation}

The eigenvalue $E^k$ corresponds to the electronic energy of the $|\Phi_{\mathrm{NOQE}}^{k}\rangle$ state. It is important to note that the operator $e^{\hat{\tau}_i}$ in $|\psi_i\rangle$ is constructed in the single-particle basis of the state $|\psi_{\mathrm{UHF}}\rangle$. Thus to measure off-diagonal elements, we formally need to transform both correlators to the same single-particle basis using a unitary orbital rotation. In this work, we absorb the unitary orbital rotation operators ($\hat{U}_{i\rightarrow1}$ and $\hat{U}_{j\rightarrow1}$) that are required to transform $|\psi_i\rangle$ and $|\psi_j\rangle$ to the same single-particle basis of reference 1, into the corresponding $e^{\hat{\tau}_i}$ and $e^{\hat{\tau}_j}$ circuits. Each reference state $|\psi_i\rangle$ encodes the occupation of $N$ spin orbitals using $N$ qubits, where each qubit represents one spin-orbital under the Jordan–Wigner or Bravyi–Kitaev encoding. Full details of these transformations and encodings can be found in the original NOQE paper \cite{baek2023say}.

Formation of the overlap matrix in the non-orthogonal basis of $M$ reference states requires evaluating $M(M-1)/2$ off-diagonal elements, and the Hamiltonian matrix requires the same number of off-diagonal elements, plus evaluation of $M$ diagonal matrix elements. In general, classical evaluation of matrix elements or operators between mean-field states dressed with the unitary coupled-cluster correlators becomes intractable, because the Baker–Campbell–Hausdorff (BCH) expansion does not truncate and becomes infinite \cite{bonfiglioli2011topics}. In contrast, it was shown in \cite{huggins2020non} that quantum computers can efficiently measure these matrix elements using a generalized Hadamard test, formally reducing the complexity scaling to polynomial dependence on the size of the system.

In the following, we use the original NOQE circuit as a concrete example to illustrate how to measure the matrices defined in Eq.~\eqref{matrices}. Building on this, we then develop a shadow-tomography-based measurement scheme that reduces both circuit width and depth, while improving the measurement efficiency in certain parameter regimes.

\subsection{Original NQOE circuit}\label{sec: original}

\begin{figure}[htbp]
    \centering
    \includegraphics[width=\linewidth]{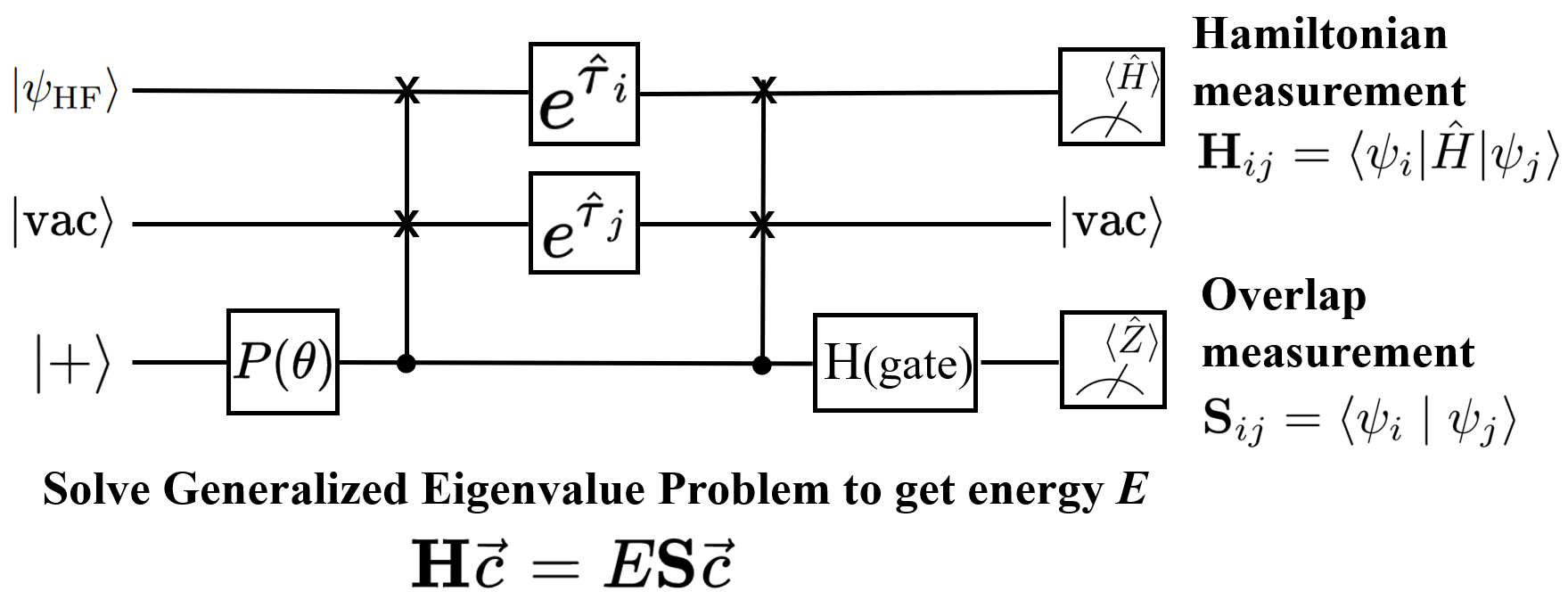}
    \caption{Circuit diagram of the original NOQE method. Two non-orthogonal reference states, $|\psi_i\rangle$ and $|\psi_j\rangle$, are prepared and combined using controlled operations to enable modified Hadamard tests that measure $H_{ij}$ and $S_{ij}$. Here, $|vac\rangle$ denotes a vacuum state (all-zeros in occupancy vector representation). Both qubit registers, $|\psi_{\mathrm{HF}}\rangle$ and $|vac\rangle$, consist of $N$ qubits, equal to the number of spin-orbitals. After measuring these matrix elements, one solves the generalized eigenvalue problem (Eq.~\ref{gep}) on a classical computer to obtain the electronic energies.}
    \label{fig:noqe_diagram}
\end{figure}

The original NOQE protocol employs a quantum circuit for a generalized Hadamard test to estimate both the overlap and Hamiltonian matrix elements. The process prepares two non-orthogonal reference states, $|\psi_i\rangle$ and $|\psi_j\rangle$ and places them into a controlled superposition. Figure~\ref{fig:noqe_diagram} illustrates the key steps of this circuit.

Just before measurement, the system is in the state (ignoring the vacuum register)
\begin{equation}
|\phi_{ij}^{\text{end}}\rangle = \frac{1}{\sqrt{2}}\bigl(|\psi_i\rangle |+\rangle + e^{i\theta} |\psi_j\rangle |-\rangle\bigr),
\end{equation}
By measuring the last qubit in the computational ($Z$) basis, one extracts either the real or imaginary part of the desired matrix element, depending on the choice of $\theta$ of the phase gate $P(\theta)$.

For $\theta=0$, measuring the last  (ancilla) qubit, yields the real part of the overlap matrix
\begin{equation}
\langle\phi_{ij}^{\text{end}}|I\otimes Z|\phi_{ij}^{\text{end}}\rangle = \mathrm{Re}(\langle \psi_i|\psi_j\rangle) = \mathrm{Re}(S_{ij}).
\end{equation}
Similarly, introducing the system Hamiltonian 
Similarly, introducing the system Hamiltonian 
\begin{equation}
    \hat{H}=\sum_{k=1}^{\Omega} w_k P_k
\end{equation}
where each $P_k$ is a Pauli operator specified by Jordan-Wigner transformation of the fermionic Hamiltonian,  allows one to measure
\begin{equation}
\langle\phi_{ij}^{\text{end}}|H\otimes Z|\phi_{ij}^{\text{end}}\rangle = \mathrm{Re}(\langle \psi_i|\hat{H}|\psi_j\rangle) = \mathrm{Re}(H_{ij}).
\end{equation}
Adjusting $\theta$ to $\pi/2$ enables extraction of the corresponding imaginary components. Full details of the measurement procedure are given in Appendix \ref{Circuit Construction}.

Once $H_{ij}$ and $S_{ij}$ are determined for all $i,j$, one solves the generalized eigenvalue problem of Eq.~\eqref{gep} classically to obtain the target energies.

Note that calculations made for the original NOQE protocol introduced in Ref.~\cite{baek2023say} were implemented using the unitary operations directly. In the current study, we explicitly design, compile, and optimize the corresponding gate-level circuits. We numerically simulate and validate the performance of the circuits and provide full implementation details in Appendix~\ref{appendix:noqe}.

\subsection{Sample complexity of the original NOQE circuit}\label{sec: sample complexity}

To quantify the measurement overhead of different NOQE schemes, here we derive the sample complexity of the original NOQE circuit in Fig. \ref{fig:noqe_diagram}. Specifically, we consider the number of circuit repetitions $n$ required to estimate a Hamiltonian matrix element to within an additive error $\epsilon$. Here, \( \epsilon \) refers to the dimensionless additive error in estimating a properly rescaled version of the observable; see Appendix~\ref{preliminiary} for details on the rescaling convention and unit consistency. This definition of $\epsilon$ applies throughout the paper.

Estimating the Hamiltonian matrix is generally more challenging than estimating the overlap matrix, so we focus on this harder case. Let $D = 2^N$ be the dimension of the $N$-qubit Hilbert space. Define $B$ as the upper bound of the squared Frobenius norm of the electronic Hamiltonian $\hat{H}$
\begin{equation}\label{eq:h_squared_bound}
    \mathrm{Tr}(\hat{H}^2) = \sum_k w_k^2 \, D \leq B.
\end{equation}
If there are $\Omega$ Pauli terms in $H$, the total number of measurements $n$ in order to estimate an individual matrix element $H_{ij}$ to additive error $\epsilon$ can be shown to be (See Appendix~\ref{sec: Original NOQE Measurement})
\begin{equation}\label{eq:NQOE_bound}
    n \geq \mathcal{O}\left(\frac{B\,\Omega}{D\,\epsilon^2}\right).
\end{equation}

Without loss of generality, \(\hat{H}\) can always be rescaled such that its spectral norm is bounded by \(\|\hat{H}\|_{\infty} \leq 1\), which implies \(B \leq D\). Additionally, it is known in many cases that \(\Omega\) scales polynomially with the number of spin-orbitals \(N\) (typically as \(N^4\)) \cite{huggins2021efficient}. Considering there are \(M^2\) matrix elements $\hat{H}_{ij}$ to evaluate, the total number of measurements required is then

\begin{equation}\label{eq:complexity_noqe_final}
    n \geq \mathcal{O}(\text{polylog}(D) \, M^2/\epsilon^2)
\end{equation}

Although Eq. \eqref{eq:complexity_noqe_final} shows favorable scaling with respect to the Hilbert space dimension \( D \), its scaling may become unfavorable when the desired error tolerance \( \epsilon \) is within chemical accuracy, which often requires \( \epsilon \leq 1/D \) for systems of intermediate size. Note that we use the dimensionless definition of \( \epsilon \) from Appendix~\ref{preliminiary}, so its numerical value can be directly compared with \( 1/D \).

Implementing this algorithm on today’s noisy quantum computers without error correction necessitates an error mitigation technique that introduces minimal overhead while remaining practically compatible with the NOQE framework. These considerations motivate the integration of shadow tomography into the NOQE framework, as we now discuss in Secs.~\ref{shadow} and \ref{sec:Applying Shadow Tomography to NOQE}.

\section{Shadow Tomography}\label{shadow}

\begin{figure*}
    \centering
    \includegraphics[width=0.8\linewidth]{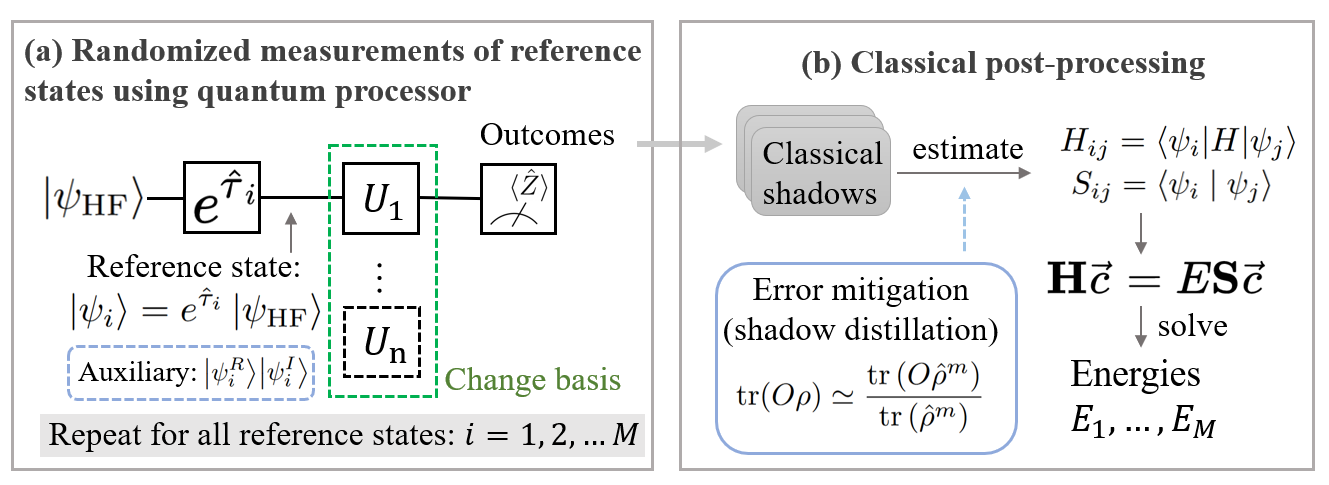}
    \caption{Overview of the shadow-tomography-enhanced NOQE protocol. (a) The UCC ansatz $e^{\hat{\tau}_i}$ is applied to the unrestricted Hartree-Fock state $|\psi_{\mathrm{HF}}\rangle$ to prepare each reference state $|\psi_i\rangle$. Randomized measurements are performed on these states and their auxiliary variants by sampling unitaries $U_1, \ldots, U_n$ from the Clifford group. This procedure is repeated for all $M$ reference states. (b) The resulting measurement outcomes are processed into classical shadows, compactly encoding the information about the reference states. From these classical shadows, one reconstructs the Hamiltonian and overlap matrix elements $H_{ij}$ and $S_{ij}$, then solves the associated generalized eigenvalue problem to obtain the energy spectrum. If needed, shadow-distillation-based error mitigation can be applied to suppress noise-induced biases in the estimated matrix elements.}
    \label{fig:main_diagram}
\end{figure*}

Shadow tomography provides a powerful and efficient method for extracting information about a quantum state from relatively few measurements \cite{huang2020predicting}. 

In Sec.~\ref{sec:shadow_data_acquisition}, we describe the data acquisition process and the construction of classical shadows. Sec.~\ref{sec:shadow_postprocess} then explains how to use these classical shadows to estimate observables.

\subsection{Data Acquisition and Classical Shadows}\label{sec:shadow_data_acquisition}

The first step in shadow tomography is to perform randomized measurements on the state of interest $\rho$. For each trial:
1. Sample a unitary $U$ at random from a suitable ensemble $\mathcal{U}$, such as the Clifford group.
2. Apply $U$ to the state $\rho$ and measure in the computational ($Z$) basis.
3. Record the outcome bitstring $b$ and the chosen unitary $U$.

From this measurement, one obtains a \emph{snapshot}
\begin{equation}
\hat{\sigma}_{U,b} = U^\dagger |b\rangle\langle b|U.
\end{equation}
Averaging $\hat{\sigma}_{U,b}$ over both $U$ and $b$ defines a measurement map $\mathcal{M}$ acting on $\rho$
\begin{equation}
\mathbb{E}_{U,b}[\hat{\sigma}_{U,b}] = \mathcal{M}(\rho).
\end{equation}

For ensembles like the Clifford group, $\mathcal{M}$ is tomographically complete and invertible. The \emph{classical shadow} of $\rho$ is defined as
\begin{equation}
\hat{\rho}_{U,b} = \mathcal{M}^{-1}(\hat{\sigma}_{U,b}),
\end{equation}
where $\mathcal{M}^{-1}$ is the reconstruction map. In particular, for $n$-qubit Clifford measurements
\begin{equation}
\mathcal{M}^{-1}(\hat{\sigma}_{U,b}) = (2^n + 1)\hat{\sigma}_{U,b} - I.
\end{equation}

It can be seen easily that the classical shadow is a sample mean estimator of the original state since one can check that 

\begin{equation}\label{eq:shadow_unbiased}
    \mathbb{E}[ \hat{\rho}_{U,b}] = \rho
\end{equation}

Since $\mathbb{E}_{U,b}[\hat{\rho}_{U,b}] = \rho$, averaging many such classical shadows yields an unbiased estimator of the state. With a sufficiently large collection of classical shadows, one can then estimate expectation values of a broad range of observables, as described next.

\subsection{Postprocessing: Predicting Properties with Classical Shadows}\label{sec:shadow_postprocess}

Once we have collected a set of classical shadows $\mathcal{D}(n) = \{\hat{\rho}_{\alpha}\}_{\alpha=1}^n$, we form an unbiased estimator $\hat{\rho} = \frac{1}{n}\sum_{\alpha=1}^n \hat{\rho}_\alpha$ for the underlying quantum state $\rho$. Given any observable $O$, we can then estimate its expectation value as
\begin{equation}
\hat{o} = \mathrm{Tr}[\hat{\rho} O],
\end{equation}
which satisfies $\mathbb{E}[\hat{o}] = \mathrm{Tr}[\rho O]$.

A key advantage of shadow tomography is its efficiency in estimating multiple observables. Specifically, under Clifford group measurements, to estimate $K$ observables $\{O_l\}_{l=1}^K$ each within additive error $\epsilon$, a sample complexity of
\begin{equation}\label{sample_complexity_1}
n \geq \mathcal{O}\left(\frac{\log(K) B}{\epsilon^2}\right),
\end{equation}
is sufficient, where $B$ is the bound of $B \geq \mathrm{Tr}[O_l^2]$ for all $l$ \cite{huang2020predicting}. This scaling improves over full tomography, which in the optimal case would still require $\mathcal{O}(4^N/\epsilon^2)$ samples, an exponential overhead.

While Eq.~\eqref{sample_complexity_1} applies to general copies of the same density matrix, in Sec.~\ref{sec: Protocol Refinement} we show that for NOQE reference states, which are pure, these bounds can be further optimized. Moreover, we will introduce specialized postprocessing techniques tailored for estimating matrix elements $H_{ij}$ and $S_{ij}$, since these are not simple linear functionals of a single reference state but bilinear forms involving pairs of states.

\section{Applying Shadow Tomography to NOQE}\label{sec:Applying Shadow Tomography to NOQE}

The original NOQE algorithm employs a modified Hadamard-test-like circuit (Fig.~\ref{fig:noqe_diagram}) to measure $H_{ij}$ and $S_{ij}$. Here, we explore the replacement of that procedure with shadow tomography measurements, thereby reducing both the circuit depth and the number of qubits required. This approach also improves the scaling of the number of measurements $n$ with respect to the number of reference states $M$, reducing it from quadratic to linear. In addition, when high precision is required for systems of moderate size, this method offers improved sample complexity compared to the original protocol. Figure~\ref{fig:main_diagram} illustrates the pipeline of our shadow-based NOQE algorithm, which consists of two main stages: data acquisition and data postprocessing.

\subsection{Data Acquisition: Preparing States and Performing Randomized Measurements}

Because the required matrix elements $H_{ij}$ and $S_{ij}$ do not correspond to simple single-state observables, we must collect data from additional states to assist in their estimation. Specifically, for each reference state $|\psi_i\rangle$, we also prepare the auxiliary states
\begin{equation}\label{eq:real_imag_ref_states}
|\psi_i^R\rangle = \frac{|0\rangle^{\otimes n} + |\psi_i\rangle}{\sqrt{2}}, \quad |\psi_i^I\rangle = \frac{|0\rangle^{\otimes n} + i|\psi_i\rangle}{\sqrt{2}}.
\end{equation}

\begin{figure}[htbp]
    \centering
    \includegraphics[width=0.9\linewidth]{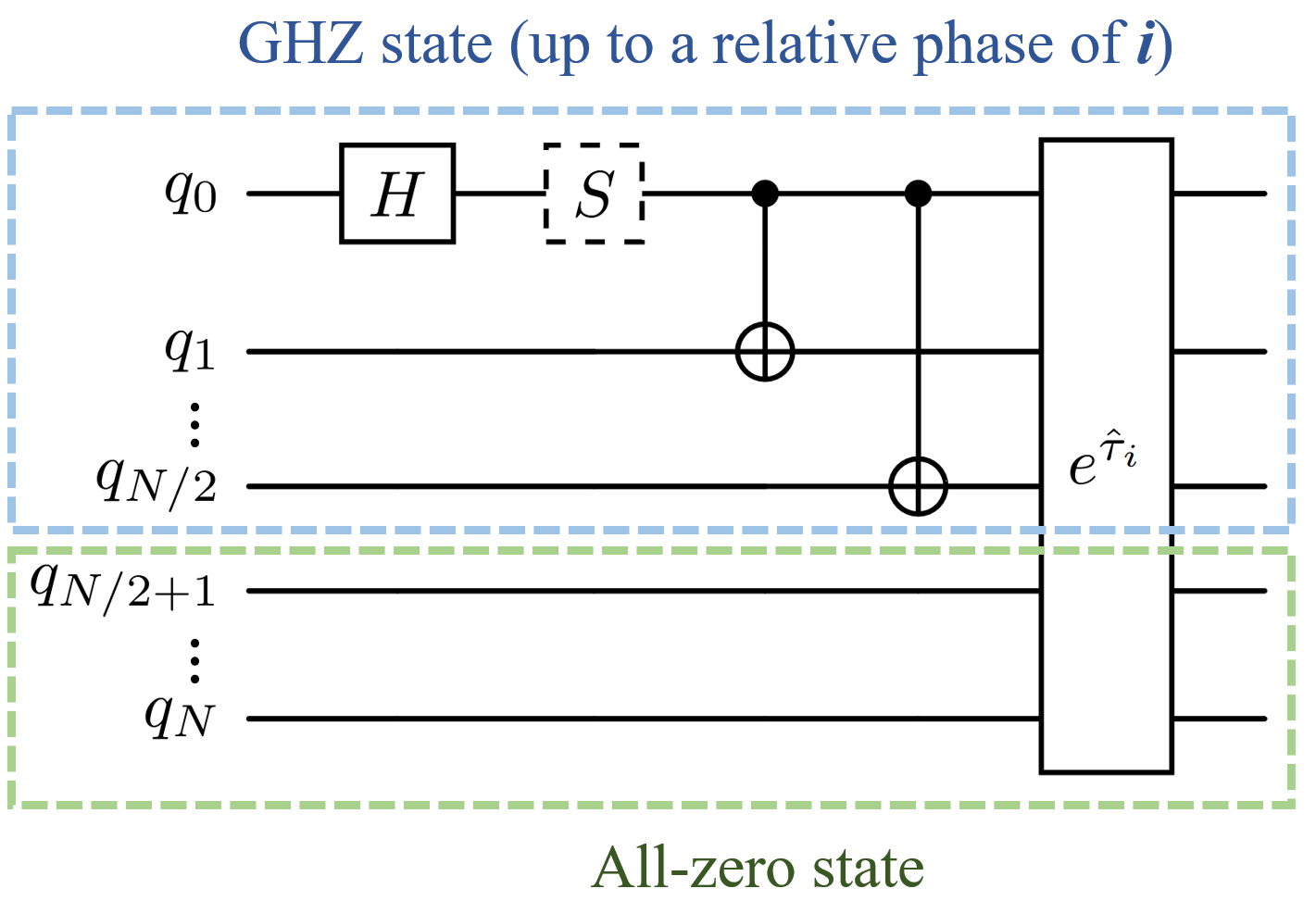}
    \caption{Quantum circuit for preparing the auxiliary states $|\psi_i^R\rangle$ and $|\psi_i^I\rangle$.}
    \label{fig:state_prep_circuit}
\end{figure}

Figure~\ref{fig:state_prep_circuit} shows an efficient circuit construction for the auxiliary states. The Hartree-Fock (HF) state for $n$ spin orbitals is $|111\cdots 000\rangle$ (i.e., $N/2$ ones followed by $N/2$ zeros) \cite{mcardle2020quantum}. Starting from an $N$-qubit zero state, we first generate an $N/2$-qubit GHZ state alongside $N/2$ qubits in the zero state. Applying the UCC ansatz $e^{\hat{\tau}_i}$ (Eq.~\ref{eq: ucc+hf}) to this superposition creates
\begin{equation}
\begin{aligned}
|\Psi_{\mathrm{end}}^i\rangle &= \frac{1}{\sqrt{2}}\bigl(e^{\hat{\tau}_i}|0\rangle^{\otimes n} + e^{\hat{\tau}_i}|1\rangle^{\otimes n/2}|0\rangle^{\otimes n/2}\bigr) \\
&= \frac{|0\rangle^{\otimes n} + |\psi_i\rangle}{\sqrt{2}} = |\psi_i^R\rangle.
\end{aligned}
\end{equation}

Here, $|\psi_i\rangle$ arises from acting $e^{\hat{\tau}_i}$ on the $|1\rangle^{\otimes N/2}|0\rangle^{\otimes N/2}$ configuration, while $|0\rangle^{\otimes N}$ remains unchanged due to the particle-conserving nature of the UCC ansatz. To obtain $|\psi_i^I\rangle$, we can add a phase gate $S$ on the first qubit, adding a factor of $i$ to the second term.

The additional overhead for preparing these auxiliary states is minimal (adding $n$ CNOT gates for the GHZ state preparation), making this procedure feasible for near-term devices. 

We then perform randomized measurements (as described in Sec.~\ref{sec:shadow_data_acquisition}) on the set $\{|\psi_i\rangle, |\psi_i^R\rangle, |\psi_i^I\rangle\}$ for each $i=1,\ldots,M$. This yields classical shadow datasets $\mathcal{D}_i(n)$, $\mathcal{D}^R_i(n)$, and $\mathcal{D}^I_i(n)$, each consisting of $n$ classical shadows.

After acquiring shadows for $|\psi_i\rangle$, $|\psi_i^R\rangle$, and $|\psi_i^I\rangle$, we use the resulting datasets to estimate the Hamiltonian and overlap matrix elements, as described in the next subsection.

\subsection{Postprocessing: Estimating Matrix Elements}\label{sec:Postprocessing: Estimating Matrix Elements}

After performing shadow tomography on each $|\psi_i\rangle$, $|\psi_i^R\rangle$, and $|\psi_i^I\rangle$, we obtain unbiased estimators of the density matrices of these states $\hat{\rho}_i = |\psi_i\rangle \langle\psi_i|$, $\hat{\rho}_i^R = |\psi^R_i\rangle \langle\psi^R_i|$, and $\hat{\rho}_i^I = |\psi^I_i\rangle \langle\psi^I_i|$ (see Eq.~\eqref{eq:shadow_unbiased}). These allow us to construct unbiased estimators for the matrix elements $H_{ij}$ and $S_{ij}$ by replacing the corresponding terms $|\psi_i\rangle \langle\psi_i|, \, |\psi^R_i\rangle \langle\psi^R_i|, \, |\psi^I_i\rangle \langle\psi^I_i|$ in the analytical formulas shown below with their shadow tomography estimators $\hat{\rho}_i, \,  \hat{\rho}_i^R, \,\hat{\rho}_i^I$.

The overlap matrix elements $S_{ij}$ in Eq.~\eqref{matrices} can be expressed in terms of the reference and auxiliary states as
\begin{equation}\label{eq:S}
    |S_{ij}|^2 = \mathrm{Tr}\bigl(|\psi_i\rangle\langle\psi_i|\,|\psi_j\rangle\langle\psi_j|\bigr), 
\end{equation} 
\begin{equation}\label{eq:Re_S}
    \operatorname{Re}(S_{ij}) = 2\,\mathrm{Tr}\bigl(|\psi_i^R\rangle\langle\psi_i^R|\,|\psi_j^R\rangle\langle\psi_j^R|\bigr) - \tfrac{1}{2}(1 + |S_{ij}|^2),
\end{equation}
\begin{equation}\label{eq:Im_S}
    \operatorname{Im}(S_{ij}) = 2\,\mathrm{Tr}\bigl(|\psi_i^I\rangle\langle\psi_i^I|\,|\psi_j^R\rangle\langle\psi_j^R|\bigr) - \tfrac{1}{2}(1 + |S_{ij}|^2).
\end{equation}
Using these relations, we first estimate $|S_{ij}|^2$ from the reference states, then determine the real and imaginary parts of $S_{ij}$ from the auxiliary states.

Once $S_{ij}$ is known, the Hamiltonian matrix elements follow from
\begin{equation}\label{eq:H}
H_{ij} S_{ij} = \mathrm{Tr}\bigl(|\psi_i\rangle\langle\psi_i|\,|\psi_j\rangle\langle\psi_j|\,H\bigr).
\end{equation}
Thus, having obtained $S_{ij}$, we can solve for $H_{ij}$ by calculating the quantity above. The derivations of Eqs.~\eqref{eq:S}--\eqref{eq:H} are detailed in Appendix~\ref{appendix:shadow}.

With all $H_{ij}$ and $S_{ij}$ estimated, we solve the generalized eigenvalue problem in Eq.~\eqref{gep} to find the energies. In essence, shadow tomography transforms NOQE into a hybrid classical-quantum protocol, where the task of estimating the required matrices through quantum circuits is delegated to classical postprocessing of the collected shadows. 

A detailed sample complexity analysis comparing this approach to the original NOQE method is presented in Sec.~\ref{sec: resource reduction analysis}. Here, we shall countinue with quantifying the sample complexity of the protocol. 

\subsection{Sample Complexity and Refinement with U-Statistics for Pure Reference States}\label{sec: Protocol Refinement}

The shadow tomography framework described in Sec. \ref{shadow} is designed for general density matrices. However, in our NOQE application, the reference states are pure. This allows us to improve the sample complexity by constructing more sophisticated estimators based on U-statistics \cite{kotz2012breakthroughs}. The U-statistics helps to produce minimum-variance unbiased estimators of the pure reference states. Note that the sample complexity in Eq.~\eqref{sample_complexity_1} corresponds to the case of U-statistics with $m = 1$.

Using $m=2$ U-statistics, we define the estimator of $\rho$ as
\begin{equation}\label{eq:u_stat_2}
    \hat{\rho}^{(U_2)} = \frac{1}{n(n-1)} \sum_{\alpha \neq \beta} \hat{\rho}_{\alpha} \hat{\rho}_{\beta}.
\end{equation}
Here, $n$ represents the number of samples, and the summation runs over all distinct pairs of samples. For pure states, $\mathbb{E}[\hat{\rho}^{(U_2)}] = \rho$, making it an unbiased estimator of $\rho$. Ref.~\cite{Grier2024sampleoptimal} shows that to estimate $K$ linear observables, i.e.,$Tr(\rho O_l)$ for $l = 1, \dots, K$ in a $D$-dimensional Hilbert space to within additive error $\epsilon$ using $m =2$ U-statistics, it suffices to have a sample size of
\begin{equation}\label{sample_complexity_2}
    n \geq \mathcal{O}\left(\log(K)\,\left\{\frac{\sqrt{BD}}{\epsilon} + \frac{1}{\epsilon^2}\right\}\right),
\end{equation}

This represents a notable improvement over the standard shadow tomography bound in Eq.~\eqref{sample_complexity_1}. Specifically, when the first term in parentheses dominates (i.e., for large $BD$ and $\epsilon$), the leading term $\frac{\sqrt{BD}}{\epsilon}$ in Eq.~\eqref{sample_complexity_2} exhibits Heisenberg-like scaling with respect to $\epsilon$, in contrast to the standard quadratic scaling. In the regime where the second term dominates (i.e., for small Hilbert space dimension $D$ or very small $\epsilon$), the second-order U-statistics approach still improves upon Eq.~\eqref{sample_complexity_1}, as it removes the dependence on the bound $B$ in the numerator, resulting in a more favorable constant sample complexity.

In this work, we find that using $m=3$ U-statistics can yield even better performance
\begin{equation}\label{eq:u_stat_3}
    \hat{\rho}^{(U_3)} = \frac{1}{n(n-1)(n-2)}\sum_{\alpha \neq \beta \neq \gamma} \hat{\rho}_\alpha \hat{\rho}_\beta \hat{\rho}_\gamma.
\end{equation}
For each observable $O_l$, we then estimate its expectation value as $\hat{o}_l = \mathrm{Tr}[O_l \hat{\rho}^{(U_3)}]$, which is the situation for estimating diagonal elements of the Hamiltonian $H_{ii}$. Following a similar derivation to that in \cite{Grier2024sampleoptimal}, we find that the sample complexity is (see Appendix \ref{sample_comp_u} for detail)
\begin{equation}\label{eq: sample_complexity_u3}
    n \geq \mathcal{O}\left(\log(K)\,\left\{ \frac{D^{2/3}B^{1/3}}{\epsilon^{2/3}} + \frac{1}{\epsilon^2}\right\}\right).
\end{equation}

This $m=3$ approach further improves sample complexity for a wide range of precision parameters $\epsilon$. While higher-order U-statistics ($m>3$) may offer additional improvements, deriving their sample complexities becomes increasingly challenging and we restrict our analysis here to $m=3$. 

\begin{figure}
    \centering
\begin{tikzpicture}[>=latex]

\draw[->] (-0.5, 0) -- (7, 0) node[below right] {$\epsilon$};

\coordinate (O) at (0, 0);
\coordinate (A) at (2, 0);
\coordinate (B) at (4, 0);
\coordinate (C) at (6.5, 0);

\foreach \pos/\label in {(O)/$0$, (A)/${1/\sqrt{BD}}$, (B)/${1/(B^{1/4}D^{1/2})}$, (C)/$1$} {
    \draw[thick] \pos -- ++(0, 0.1); 
    \node[below, scale = 0.75] at \pos {\label}; 
}

\draw[blue, thick, decorate, decoration={brace, amplitude=10pt, raise=13pt, mirror}] (O) -- (A) 
    node[midway, yshift=-31pt,  blue] {$\mathcal{O}\left(\frac{1}{\epsilon^2}\right)$};

\draw[blue, thick, decorate, decoration={brace, amplitude=10pt, raise=13pt, mirror}] (A) -- (C) 
    node[midway, yshift=-31pt,  blue] {$\mathcal{O}\left(\frac{\sqrt{BD}}{\epsilon}\right)$};


\draw[red, thick, decorate, decoration={brace, amplitude=10pt, raise=5pt}] (O) -- (B) 
    node[midway, yshift=25pt, red] {$\mathcal{O}\left(\frac{1}{\epsilon^2}\right)$};

\draw[red, thick, decorate, decoration={brace, amplitude=10pt, raise=5pt}] (B) -- (C) 
    node[midway, yshift=25pt, red] {$\mathcal{O}\left(\frac{B^{1/3}D^{2/3}}{\epsilon^{2/3}}\right)$};


\draw[blue, thick] (-0.4, 1) -- (0.1, 1) node[right, blue] {$m=2$};
\draw[red, thick] (-0.4, 0.6) -- (0.1, 0.6) node[right, red] {$m=3$};
\end{tikzpicture}

 \caption{The sample complexity of linear function shadow estimation can be optimized by using U-statistics of order \(m = 2\) (Eq.~\ref{sample_complexity_2}) and \(m = 3\)
(Eq.~\ref{eq: sample_complexity_u3}).
Note that the common factor \(\log(K)\) is omitted here for simplicity. 
When \(\epsilon\) becomes sufficiently small ($\epsilon \leq 1/\sqrt{BD}$), the sample complexity becomes $1/\epsilon^2$ and is independent of \(B\) and \(D\). Using higher-order U-statistics allows entry into this regime earlier at larger \(\epsilon\). A similar improvement due to U-statistics is observed in the sample complexity of shadow estimation for bilinear functions.}

\label{fig:axis}
\end{figure}
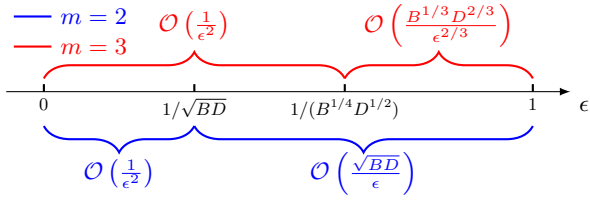

A schematic illustration of how $m=2$ U-statistics improves the original shadow estimation, and also how $m=3$ U-statistics further improves $m=2$ is shown in Fig \ref{fig:axis}. The axis is divided into several ranges of $\epsilon$. In each range, the sample complexity using different orders $m$ (denoted by different colors) of U-statistics is shown. For example, with $m=2$, if $\epsilon < 1/\sqrt{BD}$, the $1/\epsilon^2$ term in Eq. \ref{sample_complexity_2} dominates and results in a sample complexity of $\mathcal{O}(\frac{1}{\epsilon^2})$. When \(\epsilon \leq \sqrt{B/D}\), employing U-statistics with \(m = 2\) (Eq.~\ref{sample_complexity_2}) or \(m = 3\) (Eq.~\ref{eq: sample_complexity_u3}) offers an improvement over the \(m = 1\) case (Eq.~\ref{sample_complexity_1}).

For the shadow estimation of bilinear functions \(\mathrm{Tr}[O \rho_A \rho_B]\), such as the off-diagonal matrix elements \(S_{ij}\) and \(H_{ij}\), U-statistics can provide improvements similar to the linear function case. With the original bilinear shadow estimation (\(m=1\)), the sample complexity for estimating \(K\) observables to within an additive error \(\epsilon\) is given by (see Appendix \ref{Shadow Estimation of Specific Nonlinear Functions}):

\begin{equation}
n \geq \mathcal{O}\left(\mathrm{log}(K)\left\{\frac{1}{\epsilon^2} + \frac{\sqrt{D B}}{\epsilon}\right\}\right),
\end{equation}
where \(B \geq \mathrm{Tr}[\hat{H}^2]\). By employing \(m=3\) U-statistics, we demonstrate in Appendix \ref{appendix:u3_bilinear} that the sample complexity is improved to:

\begin{equation}\label{eq:samp_bilinear_u3}
    n \geq \mathcal{O}\left(\mathrm{log}(K)\left\{\frac{1}{\epsilon^2} + \frac{D^{5/6} B^{1/6}}{\epsilon^{1/3}} \right\}\right).
\end{equation}

Given the clear advantage of optimizing shadow estimations using U-statistics, we use \(m=3\) U-statistics for estimating matrix elements throughout this work.

\begin{figure}[htbp]
    \centering
    \includegraphics[width=0.5\textwidth]{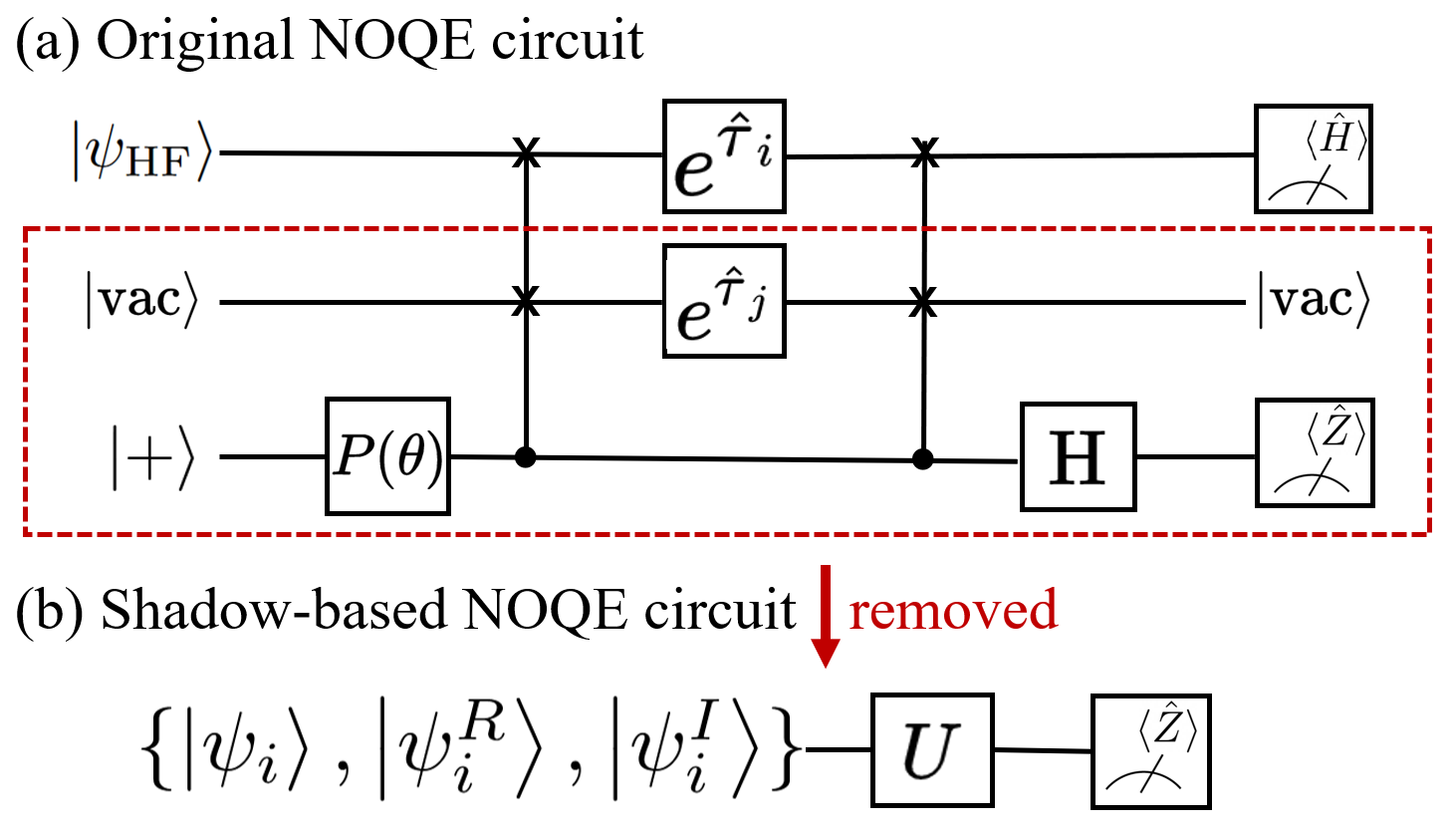}
    \caption{Comparison of the original and shadow-based NOQE measurement schemes. (a) The original NOQE circuit employs a modified Hadamard test to measure each matrix element. (b) By using shadow tomography to characterize reference states via randomized measurements and classical postprocessing, the required circuit size and number of qubits are reduced, making the implementation less demanding.}
    \label{fig:shadowNOQE}
\end{figure}

\section{Circuit Simplification and Measurements Reduction }\label{sec: resource reduction analysis}

\begin{table*}
\centering
\renewcommand{\arraystretch}{1.2}
\setlength{\tabcolsep}{3.5 pt}
\begin{tabular}{lccccc}
\toprule
 & \multicolumn{2}{c}{\textbf{Circuit Size}} 
 & \multicolumn{2}{c}{\textbf{Sample Complexity}} 
 & \textbf{Optimal Regime} \\ 
\cmidrule(r){2-3} \cmidrule(r){4-5}
 & \textbf{Qubits} & \textbf{Gates} & \textbf{\# Reference States ($M$)} & \textbf{Dimension: $D = 2^N$} & \\ 
\midrule
\textbf{Original NOQE} 
 & $2N+1$ 
 & $g$ 
 & $\mathcal{O}(M^2)$ 
 & $\mathcal{O}(\text{polylog}(D)/\epsilon^2)$, Eq.\eqref{eq:complexity_noqe_final} 
 & Larger Dimension \\ 
\textbf{Shadow-based NOQE} 
 & $N$ 
 & $\approx g/2$ 
 & $\mathcal{O}(M)$ 
 & $\mathcal{O}(1/\epsilon^2 + D/\epsilon^{1/3})$, Eq.~\eqref{eq:samp_bilinear_u3}
  & More Reference States,\\ & & & & & Higher Precision \\ 
\bottomrule
\end{tabular}
\caption{Comparison between original and shadow-based NOQE methods. The shadow-based approach reduces the number of qubits, gates, and reference state preparations, resulting in relative resource reduction and improved sample complexity in the regime of high precision $\epsilon$, a large number of reference states $M$, and a moderate number of qubits $N$.}
\label{tab:bound_comparison}
\end{table*}

\begin{figure*}[htbp]
\centering 
\begin{subfigure}{.5\textwidth}
  \centering
  \includegraphics[width=8.5cm]{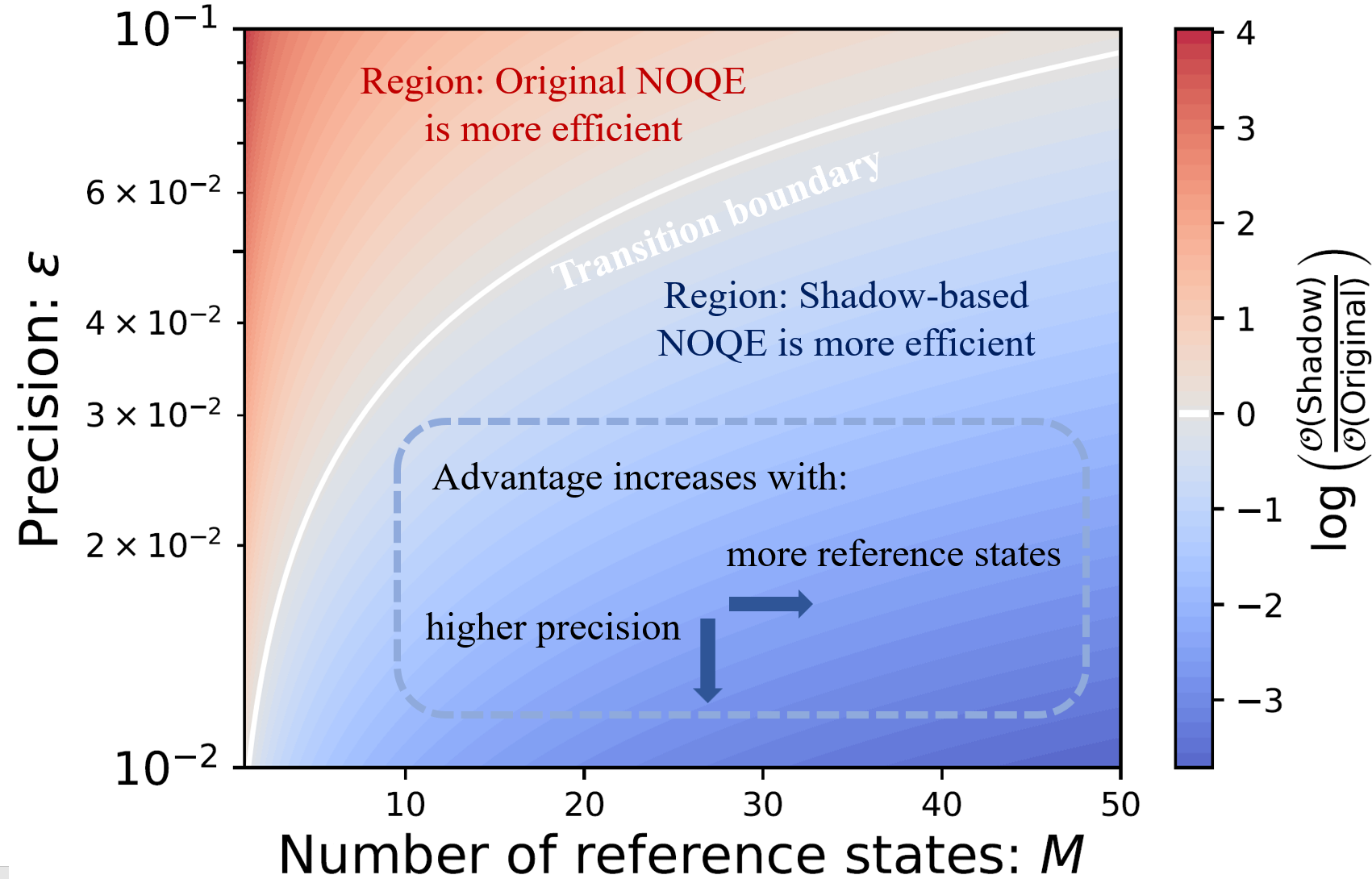}
  \caption{}
  \label{}
\end{subfigure}%
\begin{subfigure}{.5\textwidth}
  \centering
  \includegraphics[width=8.5cm]{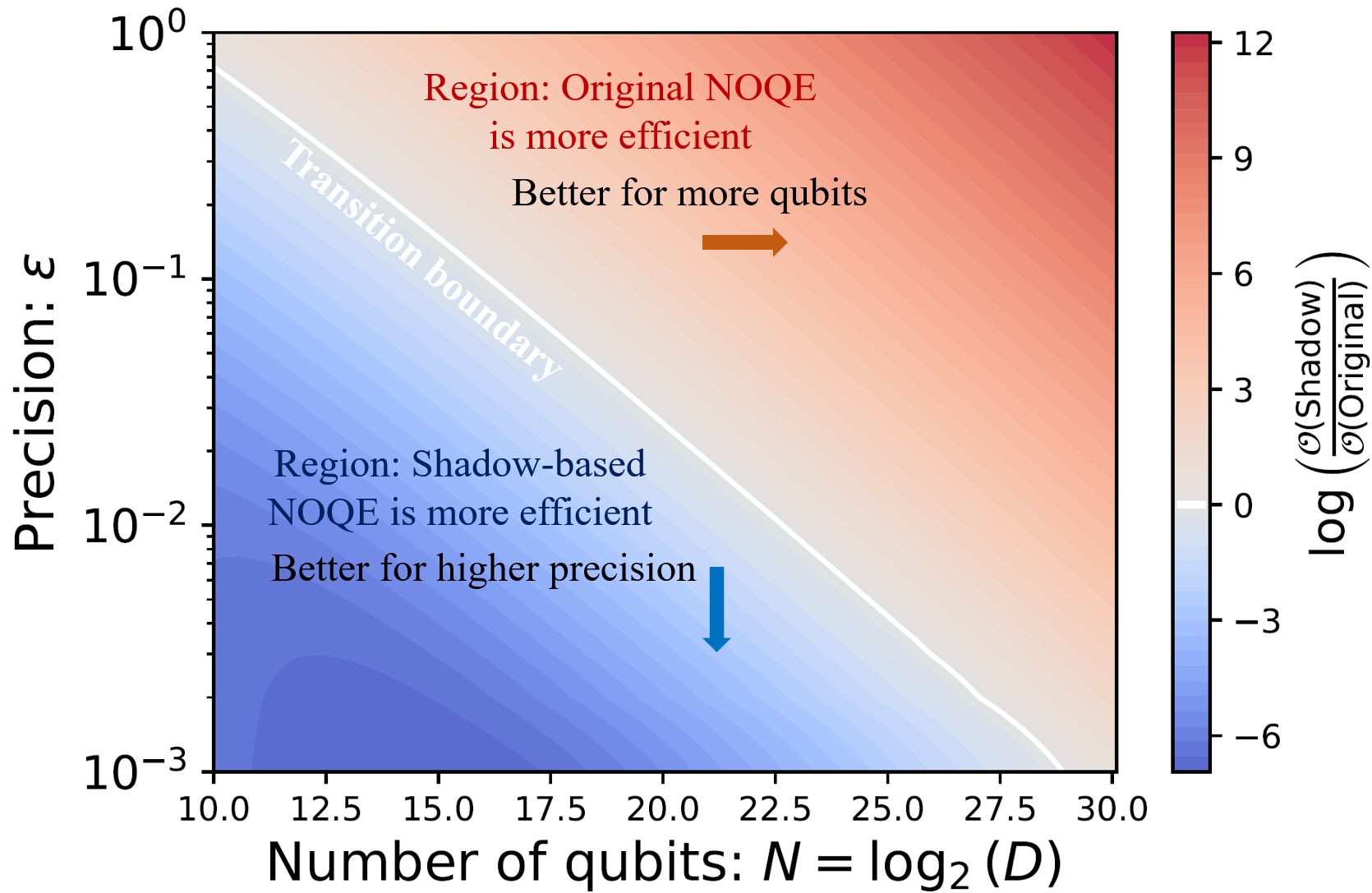}
  \caption{}
  \label{}
\end{subfigure}
\caption{Sample complexity comparison between the original and shadow-based NOQE protocols. The color scale represents the logarithmic ratio of computational cost, where red regions indicate that the original NOQE is more efficient, while blue regions indicate that the shadow-based NOQE is more efficient. (a) Varying precision $\epsilon$ and the number of reference states $M$, with the number of qubits fixed at $N = 20$. (b) Varying precision $\epsilon$ and the number of qubits $N = \log_2(D)$, with the number of reference states fixed at $M = 6$. The white contour represents the transition boundary where both methods have comparable sample complexity. The shadow-based NOQE approach exhibits advantages in the high-precision regime and when the number of reference states is large, and the original NOQE is more efficient for larger numbers of qubits beyond $N\approx 30$.}
\label{fig:bound_plot}
\end{figure*}

\begin{table}[h!]
\centering
\renewcommand{\arraystretch}{1.5}
\setlength{\tabcolsep}{12pt}
\begin{tabular}{@{}ll@{}}
\toprule
\textbf{Protocol} & \textbf{Sample Complexity} \\ \midrule
QST (independent) & $D^3 / \epsilon^2$ \\ 
QST (joint) & $D^2 / \epsilon^2$ \\ 
Original NOQE & $\text{polylog}(D) / \epsilon^2$ \\ 
Shadow-based NOQE & $1 / \epsilon^2 + D / \epsilon^{1/3}$ \\ 
\bottomrule
\end{tabular}
\caption{Sample complexity comparison of various measurement protocols. Here $D=2^N$ denotes the dimension of the $N$-qubit system. Quantum state tomography (QST) is shown for both individual (measuring a single copy at a time) and joint (measuring multiple copies simultaneously) measurement scenarios, and both NOQE protocols have better sample complexity than state-of-the-art QST methods.}
\label{table:protocols}
\end{table}

In this section, we compare the resource requirements and sample complexities of the original and shadow-based NOQE protocols. Table \ref{tab:bound_comparison} summarizes the main findings.

For an $N$-qubit system with $M$ reference states, the original NOQE protocol using the modified Hadamard test (Fig.~\ref{fig:noqe_diagram}) requires $2N+1$ qubits to measure $O(M^2)$ matrix elements. We note that there is also a variant of the NOQE Hadamard test protocol \cite{baek2023say} that reduces the qubit count to $N+1$, this does so at the cost of deeper circuits, which we do not consider here.

By contrast, the shadow-based NOQE protocol requires only $N$ qubits—halving the qubit count—and reduces the necessary reference state preparations from $O(M^2)$ to $O(M)$. This improvement arises because matrix elements are reconstructed from postprocessing classical shadows rather than directly measuring pairs of states. Consequently, the circuit is simpler, less noise-prone, and more experimentally feasible. See Fig.~\ref{fig:shadowNOQE} for illustrations.

In addition to improvements in scaling with $M$, we also examine how performance depends on $D = 2^N$. In the worst-case scenario where \(B \sim D\), the sample complexity in Eq. \eqref{eq:samp_bilinear_u3} simplifies to:
\begin{equation}\label{eq: sample complexity u3 final}
    n \geq \mathcal{O}\left( \frac{1}{\epsilon^2} + \frac{D}{\epsilon^{1/3}} \right),
\end{equation}

for estimating bilinear functions using \(m=3\) U-statistics. Notably, there is no dependence on \(K\), as we estimate a constant number of bilinear functions (\(S_{ij}\) and \(H_{ij}\)) for each pair of reference states.

Both original and shadow-based NOQE outperform standard quantum state tomography (QST) protocols with better scaling in terms of the dimention $D$, as summarized in Table \ref{table:protocols}.

We now compare the sample complexity of shadow-based NOQE with the original NOQE. Recall that the sample complexity of the original NOQE is \(\mathcal{O} \left(M^2 \, \mathrm{polylog}(D) / \epsilon^2 \right)\). In contrast, the shadow-based NOQE with \(m=3\) U-statistics requires \(\mathcal{O} \left(M \left\{1/\epsilon^2 + D/\epsilon^{1/3}\right\}\right)\) measurements. The shadow-based NOQE thus outperforms the original NOQE in the parameter regime where \(\epsilon \lesssim (M/D)^{3/5} \, \mathrm{polylog}(D)\). The sample complexity behavior is illustrated in Fig.~\ref{fig:bound_plot}, with respect to $M$ and $\epsilon$ for fixed $N$ in panel (a), and with respect to $N$ and $\epsilon$ for fixed $B$ in panel (b).

In the high-precision regime, the shadow-based complexity reduces to $\mathcal{O}(M/\epsilon^2)$, independent of $D$. This is ideal for NOQE, where high accuracy is important for capturing complex electronic correlations. For \(\epsilon \sim 10^{-3}\), 
the advantage of shadow-based NOQE persists up to systems with \(\sim 20\) qubits. However, it is worth noting that for $N$ values larger than 30, the original NOQE measurement protocol becomes more efficient, indicating that in the fault-tolerant regime with $N \gtrsim 50$ qubits, the original NOQE is preferred.

Overall, the incorporation of shadow tomography into NOQE provides a substantial resource advantage in qubits, gates, and total measurement cost in the case of small to intermediate $N$ for near-term applications. It enables the protocol to scale more favorably with the number of reference states and improves performance under demanding precision requirements.

\section{Shadow Distillation}\label{shadow distillation}

While reducing the measurement overhead is crucial to design efficient quantum algorithms, measurement itself is an useful resource in quantum computing \cite{zhang2024solving, lewalle2024optimal}. In near-term quantum hardware implementations, mitigating the effects of noise is necessary to maintain the full potential of quantum computers \cite{RevModPhys.95.045005}. The flexible measurement scheme of shadow tomography makes it well-suited for integrating classical error mitigation protocols. Here, we adopt one such technique, known as \textit{shadow distillation} \cite{PRXQuantum.4.010303}.

Consider a target state $|\psi\rangle$ that, due to experimental imperfections such as gate errors, is instead prepared as a noisy density matrix
\begin{equation}
\rho = (1 - \varepsilon)|\psi\rangle\langle\psi| + \varepsilon\rho_{\text{error}},
\end{equation}
where $\varepsilon$ characterizes the noise level and $\rho_{\text{error}}$ represents the error component that is approximately orthogonal to the target state.

A direct estimate of $\langle \psi|O|\psi \rangle$ by computing $\mathrm{Tr}(O\rho)$ is biased. Shadow distillation addresses this by considering
\begin{equation}
\langle O \rangle_{(m)} = \frac{\mathrm{Tr}(O\rho^m)}{\mathrm{Tr}(\rho^m)},
\end{equation}
for a positive integer $m$. Ref.~\cite{PRXQuantum.4.010303} shows that
\begin{equation}\label{eq:virtual_distil}
\frac{\mathrm{Tr}(O\rho^m)}{\mathrm{Tr}(\rho^m)} \approx \langle \psi|O|\psi\rangle\left[1+\mathcal{O}\left(\left(\frac{\varepsilon}{1-\varepsilon}\right)^m\right)\right].
\end{equation}

Raising $\rho$ to the $m$-th power suppresses the error component exponentially, effectively "distilling" the pure state contribution. However, increasing $m$ also enhances statistical fluctuations from finite sampling. Thus, an optimal $m$ balances noise suppression and sampling error. Our analysis (Sec.~\ref{sec: Protocol Refinement}) suggests that $m=3$ is a good practical choice.

Implementing shadow distillation with classical shadows is straightforward: we construct $\rho^m$ from the U-statistics-based estimators $\hat{\rho}^{U_m}$ of $\rho$. Since we are already employing $m=3$ U-statistics to estimate $\{\hat{\rho}_i, \hat{\rho}_i^R, \hat{\rho}_i^I\}$, we simply replace these estimators with their normalized versions $\{\hat{\rho}_i/\mathrm{Tr}[\hat{\rho}_i], \hat{\rho}_i^R/\mathrm{Tr}[\hat{\rho}_i^R], \hat{\rho}_i^I/\mathrm{Tr}[\hat{\rho}_i^I]\}$ when evaluating matrix elements in Eqs.~\eqref{eq:S}--\eqref{eq:H}. This yields error-mitigated estimates for $S_{ij}$ and $H_{ij}$. 

Because shadow distillation is a purely classical post-processing technique, it requires no additional quantum resources. The classical shadows already contain all necessary state information to apply advanced error mitigation. While we focus on shadow distillation here, other known error mitigation methods \cite{zhao2024group, hu2022logical, jnane2024quantum, ren2024error, hu2024demonstration} can be similarly integrated.

Incorporating shadow distillation into our protocol enhances the robustness of the estimated energies against noise, advancing the practical feasibility of accurate quantum chemistry simulations on near-term quantum hardware. We will evaluate the performance of the shadow distillation developed here in Sec. \ref{sec:Performance of the protocol with noise}.

\section{Validating the Correctness and Efficiency of the Protocol}\label{evaluation}

\begin{figure*}[htbp]
\centering 
\begin{subfigure}{.33\textwidth}
  \centering
  \includegraphics[width=5.5cm]{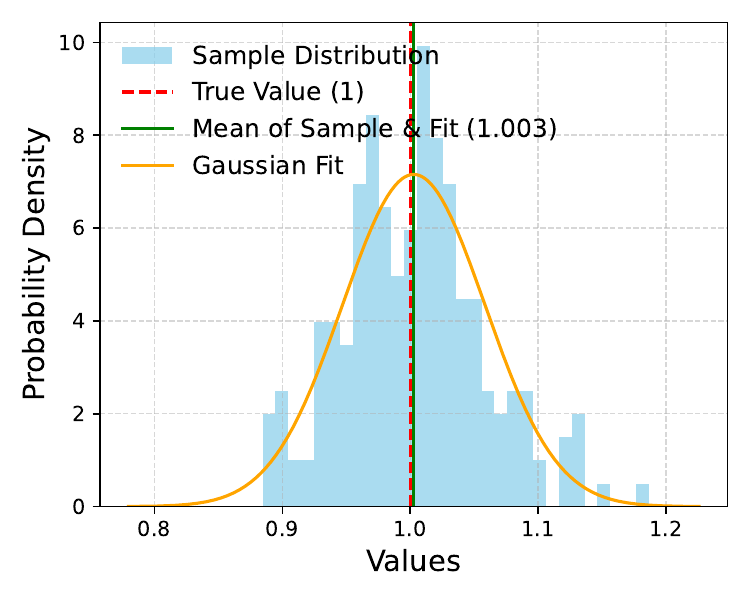}
  \caption{High-fidelity state reconstruction}
  \label{evaluation_noiseless1}
\end{subfigure}%
\begin{subfigure}{.33\textwidth}
  \centering
  \includegraphics[width=5.5cm]{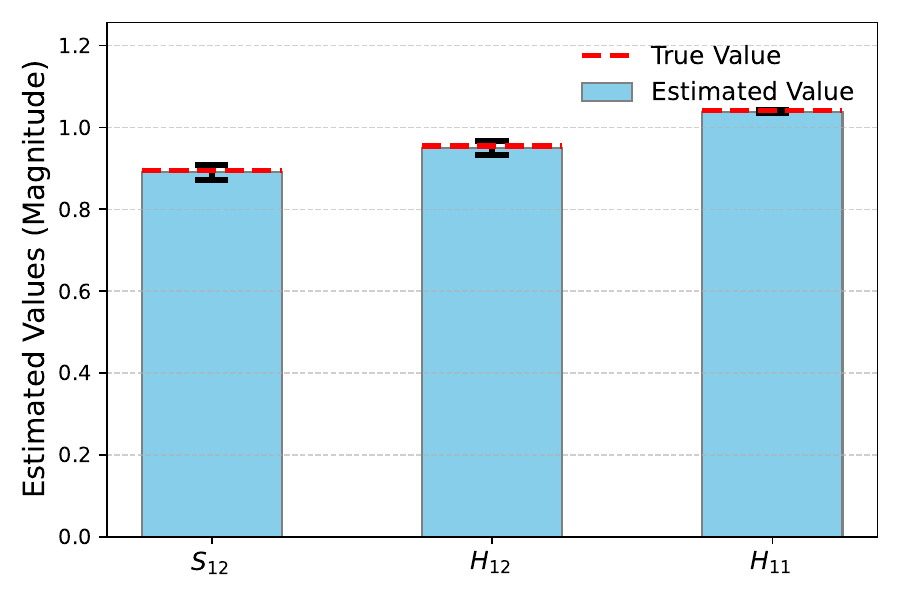}
  \caption{Accurate matrix element estimation}
  \label{evaluation_noiseless2}
\end{subfigure}
\begin{subfigure}{.33\textwidth}
  \centering
  \includegraphics[width=5.5cm]{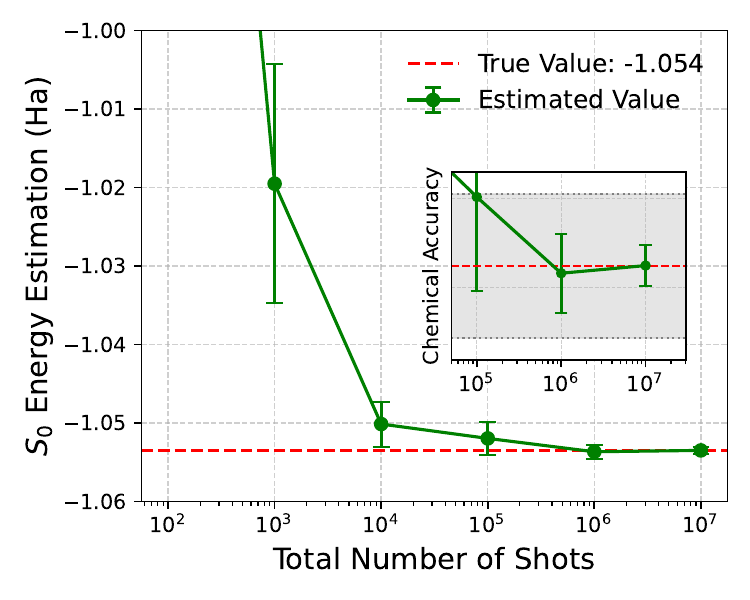}
  \caption{Energy convergence}
  \label{evaluation_noiseless3}
\end{subfigure}
\caption{Performance evaluation of the shadow-based NOQE protocol on the hydrogen molecule. (a) Fidelity distribution of the estimated reference state $|\hat{\psi}_1\rangle$ compared to the true state $|\psi_1\rangle$. Values close to unity confirm unbiased state reconstruction. (b) Histograms of estimated Hamiltonian and overlap matrix elements, showing that classical shadow postprocessing yields unbiased, precise results. (c) Convergence of the estimated ground state energy to the exact value as the number of shots increases, demonstrating the correctness and efficiency of the protocol.}
\label{evaluation_noiseless}
\end{figure*}

We validate the accuracy and efficiency of our protocol by applying it to the hydrogen molecule ($\mathrm{H}_2$) just past the Coulson-Fischer point (interatomic spacing = 1.2~\r{A}), a regime characterized by electronic spin-symmetry breaking and strong correlation \cite{coulson1949xxxiv}. This represents the most challenging case for computation and can be considered a lower bound on the performance of our protocol. 

We use the STO-3G basis \cite{hehre1969self} for a minimal orbital representation, leading to an Unrestricted Hartree-Fock (UHF) reference state $|\psi_{\mathrm{HF}}\rangle = |1100\rangle$ and two reference states $|\psi_1\rangle = e^{\hat{\tau}_1} |\psi_{\mathrm{HF}}\rangle$ and $|\psi_2\rangle = e^{\hat{\tau}_2}|\psi_{\mathrm{HF}}\rangle$. Each state encodes four spin orbitals into four qubits, i.e., $N=4$. Under these conditions, the Hamiltonian and overlap matrices have the form
\begin{equation}\label{eq: hydrogen_matrices}
\mathbf{H}=\begin{pmatrix}
h_{11} & h_{12} \\
h_{12}^* & h_{22}
\end{pmatrix}, \quad
\mathbf{S}=\begin{pmatrix}
1 & s_{12} \\
s_{12}^* & 1
\end{pmatrix}.
\end{equation}

Note that in this setting, the values to be estimated are $h_{11}$, $h_{12}$, $h_{22}$, and $s_{12}$, i.e., the number of observables $K = 4$. Solving the generalized eigenvalue problem in Eq.~\eqref{gep} provides both the ground and excited state energies. Here, we focus on the ground state energy, a primary target in quantum chemistry. We compare against the noiseless simulation results of the original NOQE protocol as a ground-truth reference (see Appendix \ref{appendix:noqe} for NOQE implementation details).

In Sec.~\ref{sec:Applying Shadow Tomography to NOQE}, we established that our protocol yields unbiased energy estimates. Figure~\ref{evaluation_noiseless} numerically confirms this. In Fig.~\ref{evaluation_noiseless1}, we assess the fidelity of reconstructed reference states. Each fidelity value is estimated from $n = 10{,}000$ classical shadows, i.e., $10{,}000$ independent runs of the circuit in Fig. \ref{fig:main_diagram}. We note that the deviation from perfect fidelity arises from statistical fluctuations, but the results remain close to unity, confirming accurate state reconstruction and reliable observable estimation.

Figure~\ref{evaluation_noiseless2} shows histograms of the estimated matrix elements. The mean and standard deviations are calculated over $100$ independent samples, each employing $10{,}000$ shadows. The estimates closely match the true values, with small error bars, indicating unbiased and precise matrix element determination.

Figure~\ref{evaluation_noiseless3} demonstrates convergence of the energy estimate to the exact value as the number of shots increases. Around $10^6$ shots (i.e., $10^6$ classical shadows) suffice to achieve chemical accuracy, defined as an absolute energy error below $1.6 \times 10^{-3}$ Hartree. Given that the exact ground state energy is $-1.054$ Hartree, this corresponds to a relative error of approximately $0.15\%$, making the estimate suitable for chemically relevant predictions \cite{wikipedia:computational_chemistry}. This level of performance—achieved with a strongly correlated system known to be challenging—highlights the protocol’s practical effectiveness.

\begin{figure}[htbp]
\centering 
\begin{subfigure}{0.5\textwidth}
  \centering
  \includegraphics[width=7cm]{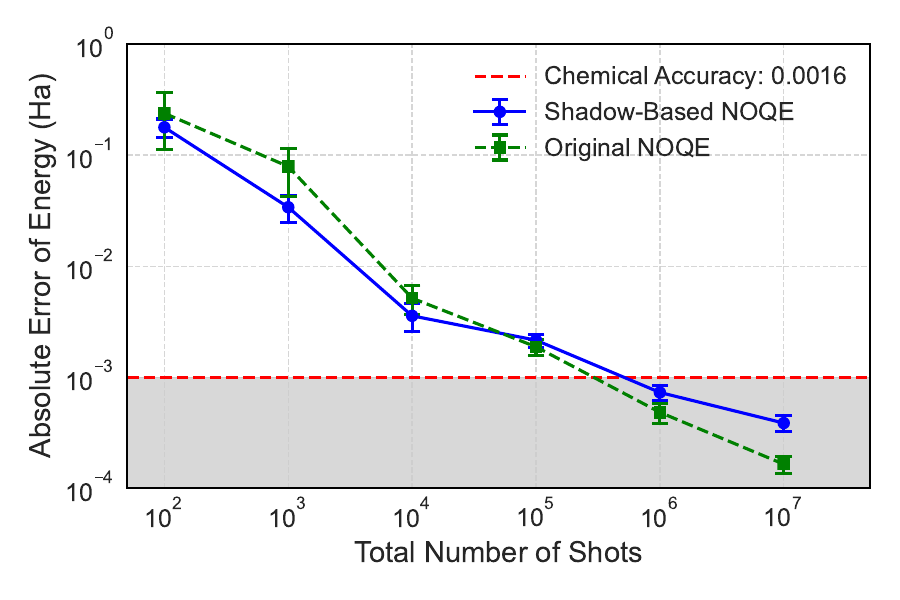}
  \caption{}
  \label{ms1}
\end{subfigure}

\begin{subfigure}{0.5\textwidth}
  \centering
  \includegraphics[width=7cm]{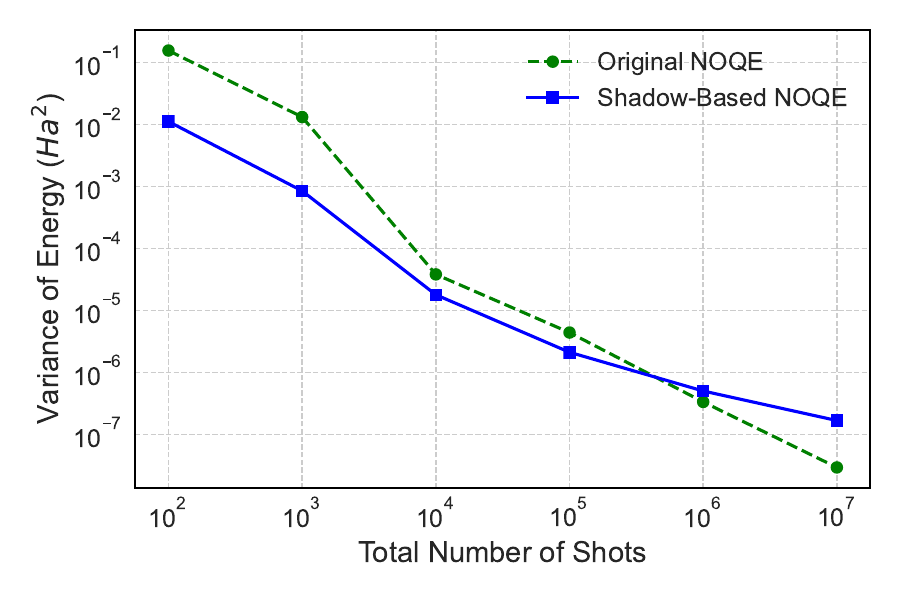}
  \caption{}
  \label{ms2}
\end{subfigure}

\caption{Comparison of (a) the absolute error and (b) the variance in energy estimation between the shadow-based NOQE protocol and the original NOQE method. Both approaches achieve chemical accuracy with a similar number of shots, but the shadow-based protocol uses half the number of qubits and gates, reducing experimental complexity and increasing feasibility for near-term implementations.}
\label{fig:absolute error of noqe and shadow}
\end{figure}

Beyond correctness, we compare circuit complexity. After circuit optimizations, the original NOQE method requires nine qubits, 124 two-qubit gates, and about 268 single-qubit gates. See Appendix \ref{Circuit resource estimation} for details. In contrast, our shadow-based approach only needs four qubits, about 54 two-qubit gates, and 110 single-qubit gates. Cutting the number of qubits and gates roughly in half significantly reduces the experimental complexity and the impact of noise.

 In terms of measurement shots, our shadow-based NOQE method, despite its reduced circuit size, performs equivalently well with the original protocol. Figure~\ref{fig:absolute error of noqe and shadow} plots the absolute error and variance in H$_2$ energy estimation versus total shots. The error bars are determined based on ten repetitions of the same experiment. Both methods reach chemical accuracy with a similar number of shots, but the shadow-based method is simpler to implement. We remark that achieving chemical accuracy for this strongly correlated four spin-orbitals system with about $10^6$ shots is on par with state-of-the-art proposals \cite{huggins2021efficient}. 

Overall, these results confirm that our shadow-tomography-enhanced NOQE protocol reliably produces accurate and precise energy estimates with fewer experimental resources, making it a promising approach for practical quantum chemistry simulations on near-term quantum devices.

\section{Performance of the Protocol under Noise}\label{sec:Performance of the protocol with noise}

In the previous sections, we showed that our estimator is unbiased in the absence of hardware noise, where the only noise present is statistical, which can be reduced by simply increasing the number of measurement shots. Real quantum devices, however, introduce systematic noise through gate errors that cannot be removed merely by taking more samples. Here, we evaluate the protocol's performance under realistic noise conditions representative of Noisy Intermediate-Scale Quantum (NISQ) devices \cite{preskill2018quantum}.

To simulate hardware noise, we use a noise model informed by the specifications of the Quantinuum H2 quantum processor \cite{quantinuum_h2_2024}. The composite noise model we used incorporates depolarizing, amplitude damping, and phase damping channels \cite{nielsen2010quantum}, calibrated to reported single-qubit and two-qubit gate error rates of $3 \times 10^{-5}$ and $1.5 \times 10^{-3}$. We introduce a scaling factor $\lambda$ to uniformly adjust all noise rates and study the protocol's behavior across varying noise strengths. The unitary operators in the NOQE algorithm were decomposed and compiled into quantum circuits consisting of single-qubit rotations and CNOT gates. Noise models, including depolarizing, amplitude damping, and phase damping channels, were applied at the level of one- and two-qubit gates to simulate realistic execution. We summarize the noise model in Tables \ref{tab:qtm_parameters} and \ref{tab:ratio_parameters} in the Appendix \ref{sec: noise model}.

\begin{figure}[htbp]
\centering 
\begin{subfigure}{0.5\textwidth}
  \centering
  \includegraphics[width=7cm]{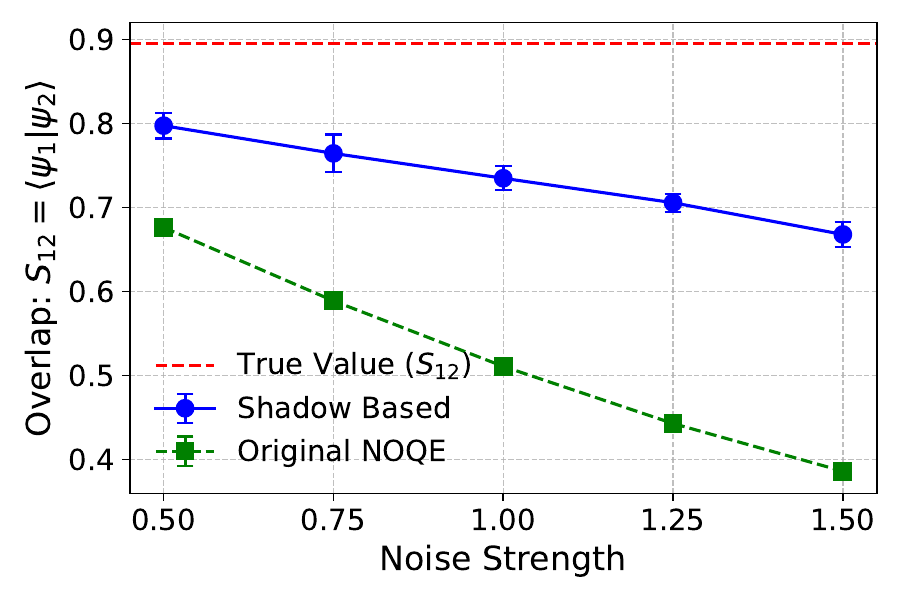}
  \caption{Overlap matrix element}
  \label{ms1}
\end{subfigure}

\begin{subfigure}{0.5\textwidth}
  \centering
  \includegraphics[width=7cm]{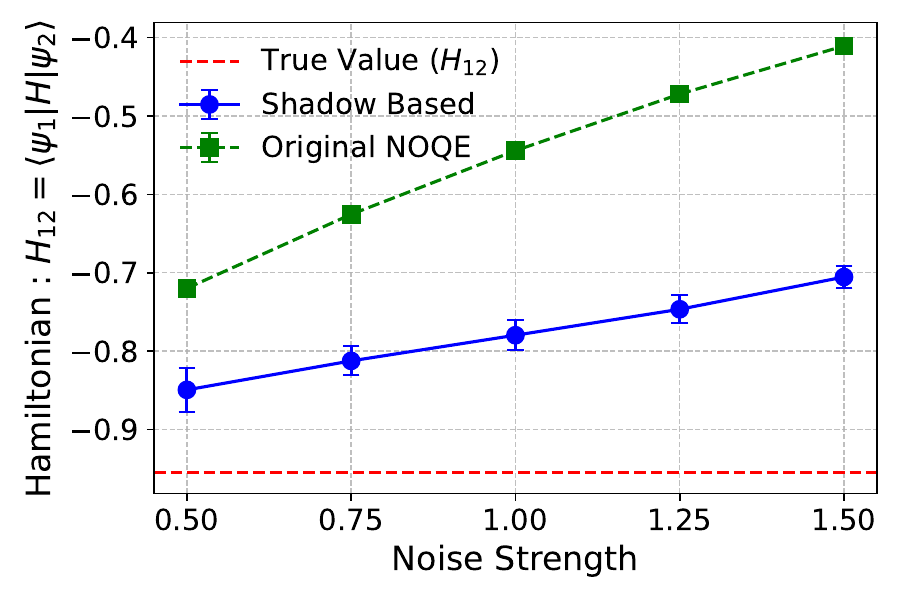}
  \caption{Hamiltonian matrix element}
  \label{ms2}
\end{subfigure}

\caption{Noise robustness of the shadow-based approach. Under noise levels scaled by factors of 0.5, 0.75, 1, 1.25, and 1.5 relative to the Quantinuum H2 error rates, we compare estimation of (a) the overlap and (b) the Hamiltonian matrix elements between the shadow-based NOQE protocol and the original NOQE method, both without error mitigations. The shadow-based approach consistently produces estimates closer to the true values, demonstrating improved resilience to hardware noise due to reduced circuit complexity}
\label{fig: noise robustness}
\end{figure}

We simulate the protocols outlined in Secs.~\ref{sec:Applying Shadow Tomography to NOQE} and \ref{shadow distillation} under these noise conditions. As shown in Fig.~\ref{fig: noise robustness}, the shadow-based protocol remains closer to the true values of both overlap and Hamiltonian off-diagonal matrix elements compared to the original NOQE method, across a range of noise strengths. This enhanced noise robustness arises partly from the simplified circuit structure: by using half as many qubits and gates, the shadow-based approach reduces error accumulation, improving overall fidelity in noisy environments.

\begin{figure}
    \centering
    \includegraphics[width=\linewidth]{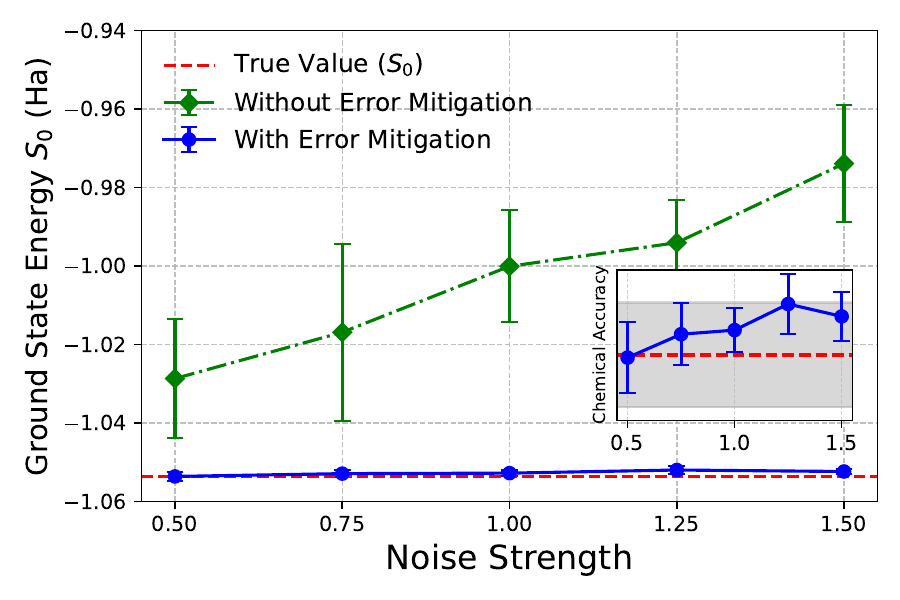}
    \caption{Energy estimation with and without error mitigation. Without mitigation (green), noise biases the energy estimates away from the true value (red). With shadow distillation (blue), the noise-induced bias is significantly removed, bringing the estimates back within chemical accuracy (gray band) across a broad range of noise strengths.}
    \label{fig:with_without_em}
\end{figure}

While the shadow-based protocol exhibits inherent noise robustness, as seen in Fig.~\ref{fig: noise robustness}, this intrinsic resilience only partially mitigates the impact of noise. To further enhance the accuracy of our protocol, we apply shadow distillation (Sec.~\ref{shadow distillation}). 

Figure~\ref{fig:with_without_em} illustrates that shadow distillation effectively suppresses noise-induced biases. Without mitigation, energy estimates drift from the true value and exhibit larger variance; with shadow distillation, they remain near the true energy within chemical accuracy. The performance remains stable across a broad range of noise regimes, demonstrating the protocol's robustness and adaptability for quantum processors with varying noise characteristics.

\begin{figure}
    \centering
    \includegraphics[width=\linewidth]{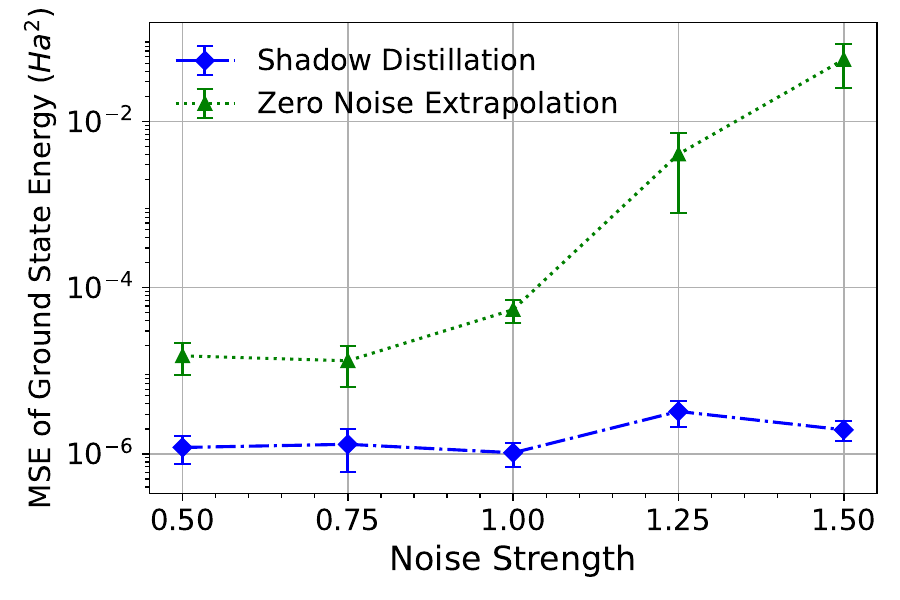}
    \caption{Mean Squared Error (MSE) comparison. The shadow-based NOQE with shadow distillation (blue) outperforms the original NOQE method with ZNE (green), achieving lower MSE and indicating better performance in reducing noise-induced bias and saving experimental resources.}
    \label{fig: mse_s0_t1_comparison}
\end{figure}

To benchmark our approach, we compare it against Zero-Noise Extrapolation (ZNE) \cite{PhysRevX.8.031027}, a widely used error mitigation method that requires additional circuits at artificially increased noise levels \cite{cheng2023fidelity}. Using the Mitiq library \cite{Mitiq}, we implement ZNE on the original NOQE method (See Appendix \ref{sec: noise model} for details) and compare its performance to the shadow-based NOQE with shadow distillation. We measure performance via the Mean Squared Error (MSE), which captures both bias and variance. Figure~\ref{fig: mse_s0_t1_comparison} shows that the shadow-based approach with shadow distillation yields lower MSE than the original NOQE with ZNE, indicating more efficient noise suppression and reduced resource requirements.

In summary, the shadow-based NOQE protocol reduces the circuit complexity, thereby inherently robust to a portion of the noise. Integrating shadow distillation further improves accuracy, effectively removing noise-induced bias. Compared to ZNE, our approach achieves lower MSE in energy estimates. This highlights the effectiveness of our protocol in noisy quantum computing environments.

\section{Summary and Outlook}\label{sec:discussion}

We have introduced a shadow-tomography-enhanced Non-Orthogonal Quantum Eigensolver (NOQE) protocol that reduces qubit requirements, circuit depth, and measurement overhead. Our rigorous sample complexity analysis shows a quadratic speedup with respect to the number of reference states, relative to the original NOQE measurement protocol. Furthermore, in the high-precision regime, the complexity no longer depends on the system size, making the protocol particularly well-suited for accurate electronic structure energy estimation on moderate-size systems. Numerical simulations of the hydrogen molecule confirm unbiased and highly accurate energy estimations.

In addition to improving sample complexity and resource requirements, the shadow-based protocol exhibits enhanced noise robustness due to its simpler circuit structure. It also naturally supports error mitigation through shadow distillation, effectively suppressing noise-induced biases without additional quantum overhead. Our results show that this error-mitigated approach outperforms common techniques like zero-noise extrapolation (ZNE), achieving chemical accuracy in estimating the hydrogen molecule's ground-state energy with as few as $10^6$ shots per reference state under realistic NISQ conditions.

While we have focused on NOQE, our methods can be extended to other problems that involve estimating overlaps or Hamiltonian matrix elements involving multiple states. If fault-tolerant quantum computing becomes accessible, our protocol could be further optimized by performing joint measurements on entangled reference states and incorporating quantum-gradient estimation algorithms \cite{huggins2022nearly, gilyen2019optimizing}.

We also note that techniques like generalized quantum subspace expansion \cite{yoshioka2022generalized} may further enhance error suppression by considering linear combinations of $\rho^m$ at multiple values of $m$ rather than a single $m$. Exploring this and other advanced error mitigation strategies within the shadow tomography framework could yield even more noise-resilient algorithms for quantum chemistry.

Future investigations into larger and more complex molecular systems, as well as experimental demonstrations on actual quantum hardware, will help further assess the scalability and practicality of our approach.

Overall, our noise-resilient, shadow-based NOQE protocol offers a promising path toward practical, high-accuracy quantum algorithms solving electronic structure problems in near-term quantum devices.

\section{Acknowledgements}
The authors thank Unpil Baek, Oskar Leimkuhler, Brandon Schramm, James Shee, Torin Stetina, and William J. Huggins for their valuable discussions. 

This work was partially supported by a joint development agreement between UC Berkeley and Dow, by the National Science Foundation (NSF) Quantum Leap Challenge Institutes (QLCI) program through Grant No. OMA-2016245, and by the U.S. Department of Energy, Office of Science, National Quantum Information Science Research Centers, Quantum Systems Accelerator. WMB is supported by the National Science Foundation Graduate Research Fellowship Program under Grant No. DGE 2146752. Any opinions, findings, and conclusions or recommendations expressed in this material are those of the authors and do not necessarily reflect the views of the National Science Foundation.

\clearpage
\newpage

\renewcommand{\thesection}{\Alph{section}}
\renewcommand{\thesubsection}{\thesection.\arabic{subsection}}

\titleformat{\section}
  [block]
  {\normalfont\Large\bfseries\centering}
  {Appendix \thesection:}
  {0.5em}
  {}

\titleformat{\subsection}
  [block]
  {\normalfont\large\bfseries\centering}
  {Appendix \thesection.\thesubsection:}
  {0.5em}
  {}

\appendix

\onecolumngrid

\section*{APPENDIX}
\addcontentsline{toc}{section}{APPENDIX}

The appendix is organized as follows: Appendix \ref{appendix:noqe} provides a detailed explanation of the working principle and implementation of the NOQE framework, which were too extensive to include in the main text. Appendix \ref{appendix:shadow} discusses the estimation of matrix using the shadow-based NOQE protocol.  Appendix \ref{sec: noise model} summarizes the noise model and the implementation of Zero Noise Extrapolation (ZNE). Appendix \ref{sec:sample_complexity} details the sample complexity derivation that occur in the paper.

\setcounter{table}{0}
\renewcommand{\thetable}{A\arabic{table}}
\setcounter{figure}{0}
\renewcommand{\thefigure}{A\arabic{figure}}

\section{NOQE Algorithm and Implementation.}\label{appendix:noqe}

\begin{figure}[htbp]
    \centering    \includegraphics[width=0.8\linewidth]{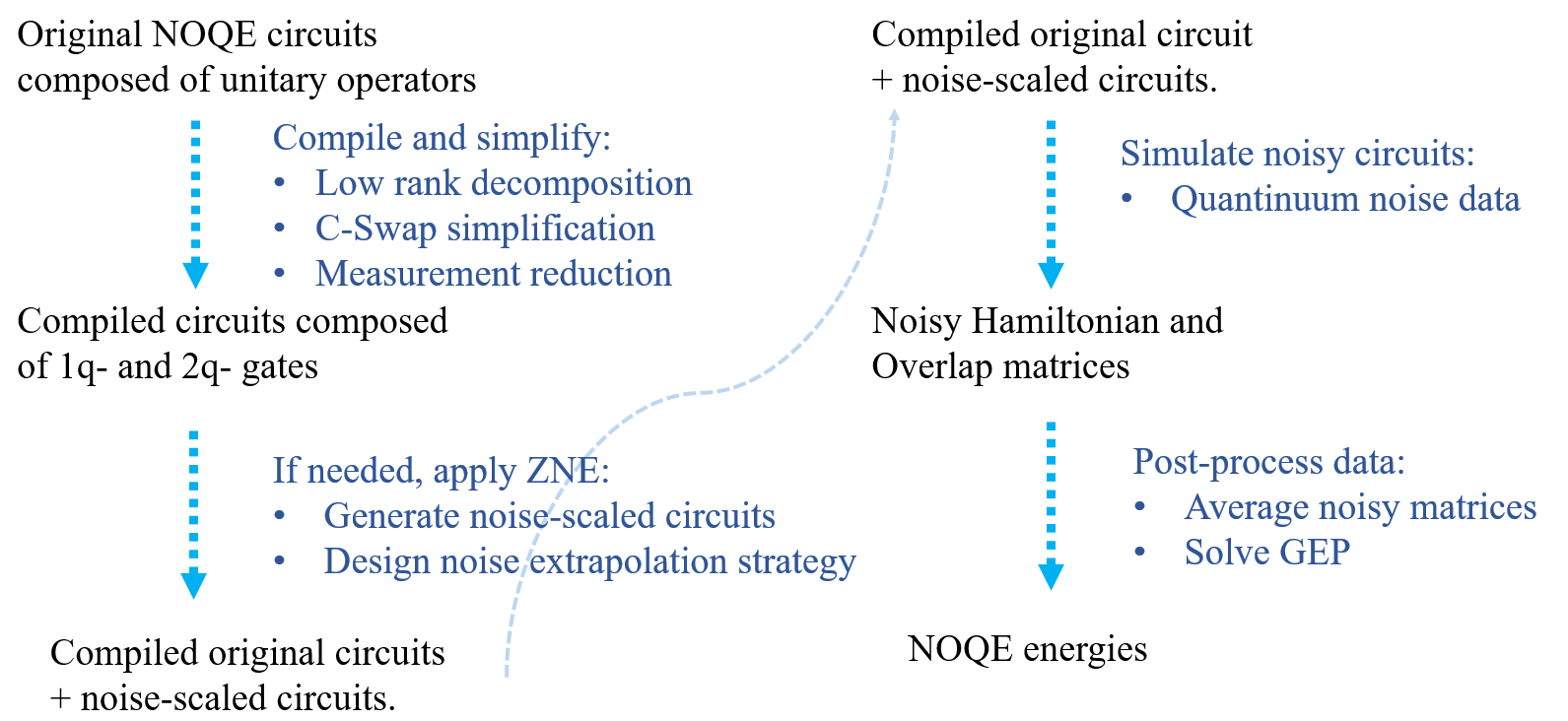}
    \caption{Working flow of implementing the NOQE algorithm}
    \label{fig:noqe_pipeline}
\end{figure}

The pipeline of implementing the original NOQE algorithm is illustrated in Fig. \ref{fig:noqe_pipeline}, with details of each step in the subsections below.

\subsection{Decomposition of Ansatzes}\label{sec:Decomposition of Ansatzes}
We use the unitary coupled-cluster doubles (UCCD) ansatz to prepare the reference state:

\begin{equation}
    \left|\psi_{\mathrm{UCCD}}\right\rangle=e^{\hat{\tau}}\left|\psi_{\mathrm{UHF}}\right\rangle
\end{equation}

$|\psi_{\mathrm{UHF}}\rangle$ is the unrestricted Hartree-Fock state, $\hat{\tau} = \hat{T}-\hat{T}^\dagger$ where

\begin{equation}
\hat{T}=\sum_{p q r s=1}^{N} t_{p s, q r} \hat{a}_{p}^{\dagger} \hat{a}_{q}^{\dagger} \hat{a}_{r} \hat{a}_{s}
\end{equation}

with  $r, s \in \operatorname{\text {occ}} ; p, q \in \operatorname{virt}$. The standard UCCD amplitudes $t_{p s, q r}$ can be approximated by the second-order many-body peturbation theory (MP2) . More specifically, we can decompose the $e^{\hat{\tau}}$ into a sum-of-squares of normal operators using singular value decomposition

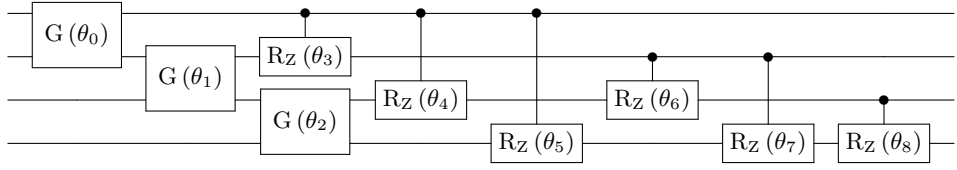
\begin{figure*}[htbp]
    \centering
    \scalebox{1}{
    \Qcircuit @C=1.0em @R=0.2em @!R { \\
	 	\nghost{  } & \lstick{  } & \multigate{1}{\mathrm{G}\,(\mathrm{\theta_0})} & \qw & \ctrl{1} & \ctrl{2} & \ctrl{3} & \qw & \qw & \qw & \qw\\
	 	\nghost{ } & \lstick{  } & \ghost{\mathrm{G}\,(\mathrm{\theta_1})} & \multigate{1}{\mathrm{G}\,(\mathrm{\theta_1})} & \gate{\mathrm{R_Z}\,(\mathrm{\theta_3})} & \qw & \qw & \ctrl{1} & \ctrl{2} & \qw & \qw \\
	 	\nghost{} & \lstick{} & \qw & \ghost{\mathrm{G}\,(\mathrm{\theta_2})} & \multigate{1}{\mathrm{G}\,(\mathrm{\theta_2})} & \gate{\mathrm{R_Z}\,(\mathrm{\theta_4})} & \qw & \gate{\mathrm{R_Z}\,(\mathrm{\theta_6})} & \qw & \ctrl{1} & \qw \\
	 	\nghost{} & \lstick{} & \qw & \qw & \ghost{\mathrm{G}\,(\mathrm{\theta_6})} & \qw & \gate{\mathrm{R_Z}\,(\mathrm{\theta_5})} & \qw & \gate{\mathrm{R_Z}\,(\mathrm{\theta_7})} & \gate{\mathrm{R_Z}\,(\mathrm{\theta_8})} & \qw  \\
\\ }}
    \caption{The basic circuit subroutine for $e^{\hat{\tau}}$ implementation. $G(\theta)$ are given rotations that realize the $\tilde{\mathcal{U}}_{B}^{(l, \mu)}$ basis rotation terms in Eq.\ref{lowrank}. Controlled-Z rotations realize the $\hat{n}_{p} \hat{n}_{q}$ number operator relevant terms. Four subsequent such subroutines realize a $e^{\hat{\tau}}$ ansatz. Single-qubit gates are omitted here.}
    \label{fig:e_hat}
\end{figure*}

\begin{equation}
e^{\hat{\tau}}=\exp \left(-i \sum_{l=1}^{L} \sum_{\mu=1}^{m} \hat{Y}_{l, \mu}{ }^{2}\right),
\end{equation}

The normal operators are then Trotterized to obtain a low-rank representation\cite{motta2021low}
\begin{equation} \label{lowrank}
    e^{\hat{\tau}} \approx \hat{\mathcal{U}}_{B}^{(1,1) \dagger} \prod_{l=1}^{L} \prod_{\mu=1}^{m} \exp \left(-i \sum_{p q}^{\rho_{l}} \lambda_{p}^{(l, \mu)} \lambda_{q}^{(l, \mu)} \hat{n}_{p} \hat{n}_{q}\right) \tilde{\mathcal{U}}_{B}^{(l, \mu)}
\end{equation}

where $\lambda_{p}^{(l, \mu)}$ are eigenvalues of the $\hat{Y}_{l, \mu}$ operators. $\tilde{\mathcal{U}}_{B}^{(l, \mu)}$ are sequences of neighboring basis rotation. The circuit implementation of ansatz is illustrated in Fig. \ref{fig:e_hat}. The details of derivation can be found in the orginal NOQE paper. \cite{baek2023say}

\subsection{\label{Circuit Construction} Circuit Construction}

\begin{figure}[htbp]
    \centering
    \includegraphics[width=0.6\linewidth]{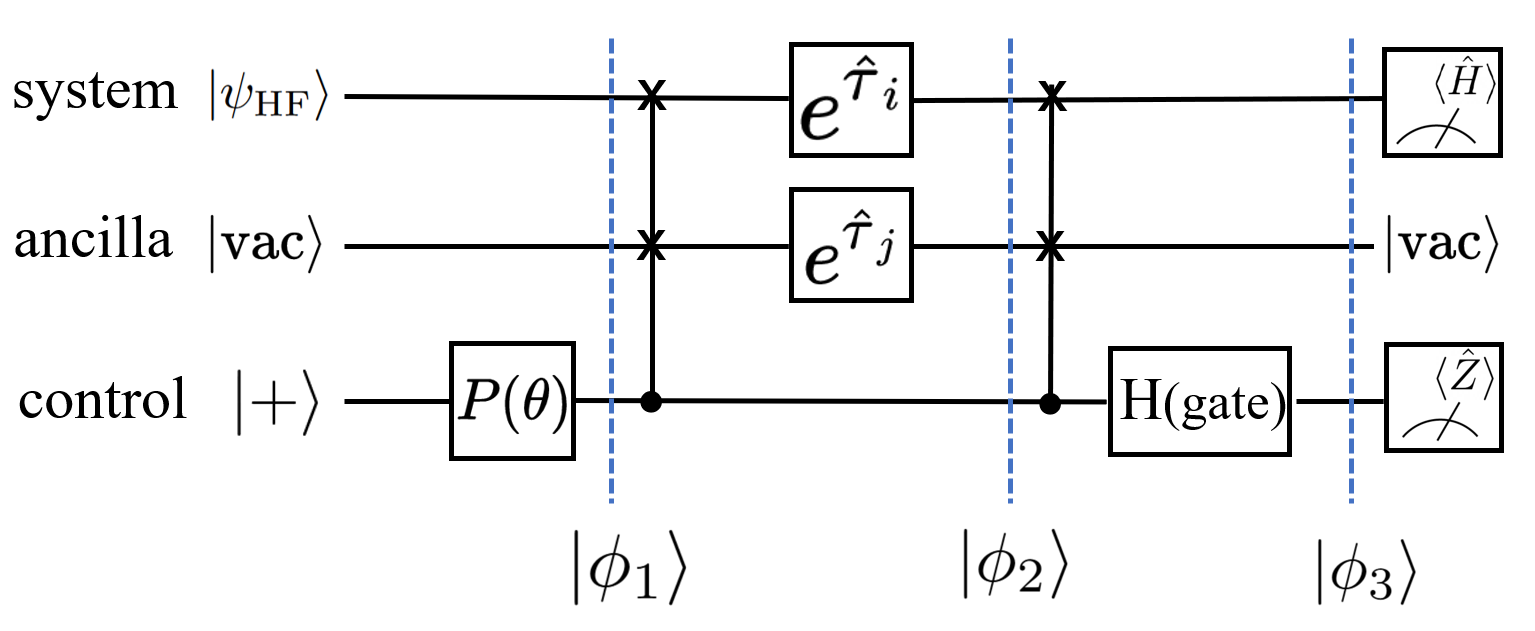}
    \caption{Original-NOQE Circuit. The quantum states in different stage of the circuits are marked and written in Eq. \ref{eq:1}, Eq. \ref{eq:2} and Eq. \ref{eq:3} }
    \label{fig:noqe_circuit_appendix}
\end{figure}

The NOQE circuit is shown in Fig.\ref{fig:noqe_circuit_appendix}. We will introduce how to estimate the state and energy of the H2 system under the STO-3G basis from the circuit. 

The circuit starts with system qubits in the the HF state, and ancilla qubits in the all-zero sate, and an extra control qubit in the plus state. To begin, a phase gate is applied to the control qubit, to create state $|\phi_1\rangle$

\begin{equation}\label{sec:eq1}
\left|\phi_1\right\rangle=\frac{1}{\sqrt{2}}\left(\left|\psi_{\mathrm{HF}}\right\rangle|\mathrm{vac}\rangle|0\rangle+e^{i \theta}\left|\psi_{\mathrm{HF}}\right\rangle|\mathrm{vac}\rangle|1\rangle\right)
\end{equation}

Then a Cswap and UCCD operator are applied to the HF state and produce two different reference states,
One reference state is in the system register, and the other is in the ancilla register. As shown by $|\phi_2\rangle$

\begin{equation}\label{sec:eq2}
\left|\phi_2\right\rangle=\frac{1}{\sqrt{2}}\left(\left|\psi_i\right\rangle|\mathrm{vac}\rangle|0\rangle+e^{i \theta}|\mathrm{vac}\rangle\left|\psi_j\right\rangle|1\rangle\right)
\end{equation}

Next, another CSwap is applied such that system qubits contain the information of both reference states, and we have the final state $|\phi_3\rangle$

\begin{equation}\label{sec:eq3}
\left|\phi_3\right\rangle=\frac{1}{\sqrt{2}}\left(\left|\psi_i\right\rangle|+\rangle+e^{i \theta}\left|\psi_j\right\rangle|-\rangle\right)
\end{equation}

The rest is about measurement. We can check that if the angle of the phase gate is zero, by measuring the control qubit under Z basis, we get the real part of the overlap matrix 

\begin{equation}
\left\langle\phi_3\right| I \otimes Z\left|\phi_3\right\rangle=\operatorname{Re}\left\langle\psi_i \mid \psi_j\right\rangle
\end{equation}

Similarly, if we measure the Hamiltonian observable of system qubit and Z of the control qubit, we get the real parts of the Hamiltonian matrix 

\begin{equation}
\left\langle\phi_3\right| H \otimes Z\left|\phi_3\right\rangle=\operatorname{Re}\left\langle\psi_i\right| H\left|\psi_j\right\rangle
\end{equation}

If $\theta$ is $\pi/2$, we can get imaginary parts. Once we get H and S matrices,

\begin{equation} \label{hs_matrix}
H_{ij} = \langle \psi_i|H|\psi_j\rangle,
S_{ij} = \langle \psi_i|\psi_j\rangle
\end{equation}

Then we can solve the generalized eigenvalue problem

\begin{equation} \label{generalized_eigenvalue_problem}
\mathbf{H} \vec{c}=\mathbf{E S} \vec{c}
\end{equation}

to get the state and the energy of the electronic system. These are basically the theory of the NOQE algorithm. 

\subsection{Circuit Simplification} \label{simplification}

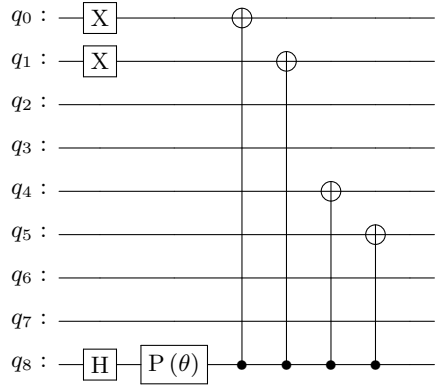
\begin{figure}[htbp]
    \centering
    \scalebox{1.0}{
    \Qcircuit @C=1.0em @R=0.2em @!R { \\
	 	\nghost{{q}_{0} :  } & \lstick{{q}_{0} :  } & \gate{\mathrm{X}} & \qw & \targ & \qw & \qw & \qw & \qw & \qw\\
	 	\nghost{{q}_{1} :  } & \lstick{{q}_{1} :  } & \gate{\mathrm{X}} & \qw & \qw & \targ & \qw & \qw & \qw & \qw\\
	 	\nghost{{q}_{2} :  } & \lstick{{q}_{2} :  } & \qw & \qw & \qw & \qw & \qw & \qw & \qw & \qw\\
	 	\nghost{{q}_{3} :  } & \lstick{{q}_{3} :  } & \qw & \qw & \qw & \qw & \qw & \qw & \qw & \qw\\
	 	\nghost{{q}_{4} :  } & \lstick{{q}_{4} :  } & \qw & \qw & \qw & \qw & \targ & \qw & \qw & \qw\\
	 	\nghost{{q}_{5} :  } & \lstick{{q}_{5} :  } & \qw & \qw & \qw & \qw & \qw & \targ & \qw & \qw\\
	 	\nghost{{q}_{6} :  } & \lstick{{q}_{6} :  } & \qw & \qw & \qw & \qw & \qw & \qw & \qw & \qw\\
	 	\nghost{{q}_{7} :  } & \lstick{{q}_{7} :  } & \qw & \qw & \qw & \qw & \qw & \qw & \qw & \qw\\
	 	\nghost{{q}_{8} :  } & \lstick{{q}_{8} :  } & \gate{\mathrm{H}} & \gate{\mathrm{P}\,(\mathrm{\theta})} & \ctrl{-8} & \ctrl{-7} & \ctrl{-4} & \ctrl{-3} & \qw & \qw\\
\\ }}
    \caption{Simplified implementation of the first Controlled-Swap network. }
    \label{fig:cswap_simplification}
\end{figure}

We will reduce the experimental resources of running the NOQE circuit in Fig.\ref{fig:noqe_diagram} by simplifying the Controlled-Swap network. The function of the first Controlled-Swap network is to prepare the state.
Considering that $|\Phi_{\mathrm{UHF}}\rangle=|1100\rangle$ and $|\operatorname{vac}\rangle=|0000\rangle$, we can design the circuit shown in Fig.~\ref{fig:cswap_simplification} to equivalently prepare the desired state of
\begin{equation}
    \frac{1}{\sqrt{2}}(|1100\rangle|0000\rangle|0\rangle+|0000\rangle|1100\rangle|1\rangle)
\end{equation}
which uses fewer two-qubit gates.

\subsection{Efficient Hamiltonian measurement} \label{h_measure}

To perform the Hamiltonian measurement, we first apply Jordan-Wigner transformation on the electronic Hamiltonian to get the corresponding qubit Hamiltonian and then decompose the it to Pauli basis, measure the expectation value of these Pauli operators and then sum them up based on the coefficient of the decomposition

\begin{equation}
H_{IJ} = \sum_{i}w_i\langle \phi_I|P_i|\phi_J\rangle
\end{equation}

Specifically, the Hamiltonian of H2 molecule under the STO-3G basis can be decomposed into 27 Pauli terms. Since the UCC-MP2 ansatz involves only double-excitation, the first reference state will be in a neat form

\begin{equation}
    |\phi_{I}\rangle = a_1|1100\rangle+a_2|0011\rangle
\end{equation}

and the second reference state is
\begin{equation}
    |\phi_{J}\rangle = a_3|1100\rangle+a_4|0011\rangle+a_5|0110\rangle+a_6|1001\rangle
\end{equation}

So we can reduce some redundant Pauli terms that produce the expectation value, i.e., 

\begin{equation} \label{pauli_equivalent}
   \langle \phi_I| P_i|\phi_J\rangle = \pm(i)\langle \phi_I| P_j|\phi_J\rangle,  P_i, P_j \in \{P_{group}\}
\end{equation}

The 27 Pauli terms can be categorized into five sets $\{P_{group}\}$: 
$\{IIII, IIZZ, IZIZ, ZIIZ, IZZI, ZIZI, ZZII\},\{ZXZX,$
$IXIX, ZYZY, IYIY, XZXZ, YZYZ, XIXI, YIYI\},$
$\{IIIZ, IIZI, IZII, ZIII\}, \{YYXX, XYYX, YXXY,$
$XXYY\}, \{IXZX, IYZY, XZZI, YZYI\}$.

It can be verified that Pauli terms belonging to the same set produce the same result (up to a $\pm$ sign and $i$), as shown in Eq.\ref{pauli_equivalent}. So the expectation values for five Pauli terms are enough to compose the Hamiltonian measurement. 

For special cases of measuring the diagonal element $\langle\phi_I|P_i|\phi_I\rangle$, the first set always produces $\pm 1$, the second and the last sets always produce $0$, so only two measurements corresponding to the third and fourth sets are necessary to be performed.

\subsection{Circuit resource estimation} \label{Circuit resource estimation}
After the simplification, in this session, we will count how many gates and circuits are needed to perform the NOQE. 

Implementing a CSwap operation requires 7 2q native gates \cite{PhysRevA.53.2855}. The number of 2q gates for the first simplified CSwap and the second complete CSwap Network is thus

\begin{equation}
    N_{CSwap}^{2q} = 4 + 4\times 7 = 32
\end{equation}

Implementing one ansatz $e^{\hat{\tau}}$ needs four subroutines in Fig.\ref{fig:e_hat}. One subroutine contains three given rotations and six CRZ rotations. Compilation of a given rotation needs two 2q gates, and CRZ needs one 2q gate.
Therefore the number of 2q gates for realizing $e^{\hat{\tau}}$  is

\begin{equation}
     N_{e^{\hat{\tau}}}^{2q} = N_{Given}^{2q}+N_{CRZ}^{2q} = 2\times4\times(3\times2+6) = 96
\end{equation}

The overall basis rotation $\hat{U}_{I\rightarrow 1}$ requires additional three given rotations,

\begin{equation}
     N_{\hat{U}_{I/J \rightarrow 1}}^{2q} = 2\times(3\times2) = 12
\end{equation}

We use Qiskit \cite{Qiskit} compiler to optimize the circuit further and count the number of 2q native gates (Rxx) and 1q rotations (U3). Here we would like to remark that the arbitrary 1q rotation can be realized by one Ry gate since U3 can be ZYZ decomposed and Rz can be virtually implemented\cite{mckay2017efficient}. With this in mind, we present the 2q- and 1q- gate numbers

\begin{equation}
     N_{all}^{2q} = 124 \quad , \quad N_{all}^{1q} = 268
\end{equation}

Above is the gate number estimation. Then we will discuss the circuit number. To evaluate the off-diagonal element of the overlap matrix, $\langle I \otimes Z\rangle$ is measured, and one circuit is needed. For the off-diagonal element of the Hamiltonian matrix, we will measure the expectation values of five Pauli operators to get $\langle H \otimes Z\rangle$, so five circuits are needed. For the diagonal element of the Hamiltonian matrix, four qubits are enough and two circuits are needed. (See \ref{h_measure} for details)

We would like to remark that We are optimizing the circuit at the gate level. Given specific device, it is possible to do circuit synthesis and optimization from pulse level which potentially allows further reduction in circuit running time.\cite{liang2023towards, cheng2024epoc}

To summarize, we will run six 9q-circuits with around 100 2q gates and two 4q-circuits with around 50 2q gates for the original NOQE circuit.

\section{Shadow-based NOQE Algorithm}\label{appendix:shadow}
Here, we detail the shadow estimation of the matrix elements used in the NOQE algorithm.

The data acquisition phase of shadow tomography performed on copies of $\rho_i$ and $\rho_j$ will eventually produces unbiased estimators $\hat{\rho}_i$ and $\hat{\rho}_j$ for these two states: $\mathbb{E}[\hat{\rho}_i] = \rho_i$ and $\mathbb{E}[\hat{\rho}_j] = \rho_j$. They are then used to form unbiased estimators $\mathrm{Tr}(\hat{\rho}_i \hat{\rho}_j O)$ for the bilinear functions $\mathrm{Tr}(\rho_i \rho_j O)$ of interest. 

In particular, in order to estimate the overlap matrix elements $S_{ij}$, such bilinear functions include 
$$O = I, \; \text{with}\; \rho_{i} = |\psi_{i}\rangle \langle \psi_{i}|,\; \rho_{i} = |\psi_{i}^{R}\rangle \langle \psi_{i}^{R}|\;\; \text{or} \;\;\rho_{i} = |\psi_{i}^{I}\rangle \langle \psi_{i}^{I}|, \; \text{while} \; \rho_j = |\psi_{j}\rangle \langle \psi_{j}| \; \text{or} \; \rho_j = |\psi_{j}^R\rangle \langle \psi_{j}^R| ,$$
where the auxiliary states $|\psi_i^{R}\rangle$ and  $|\psi_i^{I}\rangle$ are defined in Eq. \eqref{eq:real_imag_ref_states} and $|\psi_{i/j}\rangle$ are the reference states. $\mathrm{Re}(S_{ij})$ and $\mathrm{Im}(S_{ij})$ can then be extracted following Eq. \eqref{eq:S} - Eq. \eqref{eq:Im_S}. 

To prove Eq. \eqref{eq:Re_S} and Eq. \eqref{eq:Im_S}, we notice that $\langle 0|\psi_i\rangle = 0$ for any reference states $|\psi_i\rangle$. Therefore, we have for $\mathrm{Re}(S_{ij})$
\begin{equation}\label{eq: real_overlap}
\begin{aligned}
& \operatorname{Tr}\left(\left|\psi_i^I\right\rangle\left\langle\psi_i^I||\psi_j^I\right\rangle\left\langle\psi_j^I\right|\right). \\
&=\frac{1}{4} \operatorname{Tr}[\left(|0\rangle\langle 0|+| 0\rangle\left\langle\psi_i|+| \psi_i\right\rangle\left\langle 0|+| \psi_i\right\rangle\left\langle\psi_i\right|\right)
\left(|0\rangle\langle 0|+| 0\rangle\left\langle\psi_j|+| \psi_j\right\rangle\left\langle 0|+| \psi_j\right\rangle\left\langle\psi_j\right|\right)] \\
&=\frac{1}{4} \operatorname{Tr}\left[|0\rangle\langle 0|+| 0\rangle\left\langle\psi_j|+| 0\right\rangle\left\langle\psi_i \mid \psi_j\right\rangle\left\langle 0|+| \psi_i\right\rangle\left\langle 0|+| \psi_i\right\rangle\left\langle\psi_j|+| \psi_i\right\rangle\left\langle\psi_i \mid \psi_j\right\rangle\left\langle 0|+| \psi_i\right\rangle\left\langle\psi_i \mid \psi_j\right\rangle\left\langle\psi_j\right|\right]. \\
&= \frac{1}{4}(1+\left\langle\psi_i \mid \psi_j\right\rangle+\left\langle\psi_j \mid \psi_i\right\rangle+\left|\left\langle\psi_i \mid \psi_j\right\rangle\right|^2)=\frac{1}{4}\left[1+2 \operatorname{Re}\left(S_{i j}\right)+\left|S_{i j}\right|^2\right].
\end{aligned}
\end{equation}
And for $\mathrm{Im}(S_{ij})$ we have 
\begin{equation}\label{eq:imaginary_overlap}
\begin{aligned}
\operatorname{Tr}\left(\left|\psi_i^I\right\rangle\left\langle\psi_i^I|| \psi_j^R\right\rangle\left\langle\psi_j^R\right|\right) & =\frac{1}{4}[1-i\left\langle\psi_i \mid \psi_j\right\rangle+ i\langle \psi_j\left|\psi_i\right\rangle+\left|\langle\psi_i\right| \psi_j\rangle|^2] \\
& =\frac{1}{4}\left[1+2 \operatorname{Im}\left(S_{i j}\right)+\left|S_{i j}\right|^2\right].
\end{aligned}
\end{equation}

In order to obtain the Hamiltonian matrix elements, the bilinear functions we need to estimate through shadow tomography include
$$O = H, \; \text{with}\; \rho_{i} = |\psi_{i}\rangle \langle \psi_{i}|\; \text{while} \; \rho_j = |\psi_{j}\rangle \langle \psi_{j}|.$$
Then with Eq. \eqref{eq:H} and $S_{ij}$ we can extract $H_{ij}$.

Finally, in order to improve the sample complexity and also reduce noise-induced bias via shadow distillation, in all the numerical simulation presented in this work we constructed the unbiased estimators $\hat{\rho}_i$ and $\hat{\rho}_j$ using order-3 U-statistics
\begin{equation}
    \hat{\rho}_i = \hat{\rho}_i^{U_3} = \frac{1}{n(n-1)(n-2)}\sum_{\alpha \neq \beta \neq \gamma}\hat{\rho}_{i,\alpha} \hat{\rho}_{i,\beta} \hat{\rho}_{i,\gamma},
\end{equation}
where $\hat{\rho}_{i,x}$ ($x = \alpha, \beta, or\, \gamma$) are classical shadows constructed from the state $\rho_i$ and $n$ is the number of classical shadows 

\section{Noise Model and Zero Noise Extrapolation}\label{sec: noise model}

In this section, we introduce the noise model used in our simulation and the zero noise extrapolation techniques.

Characterizing noise in experiments can be challenging \cite{ren2019modeling, shaffer2023sample}. To model realistic noise, we use the calibrated noise rates from the Quantinuum System H2 hardware in our simulation. The specifications of the Quantinuum System H2 are listed in Table \ref{tab:qtm_parameters}. We incorporate typical single-qubit and two-qubit gate infidelities and, to evaluate performance across different noise regimes, we rescale the gate error rates as shown in Table \ref{tab:ratio_parameters}.

\begin{table}[htbp]
    \centering
    \begin{tabular}{|l|c|c|c|}
        \hline
        \multicolumn{4}{|l|}{\textbf{General}} \\
        \hline
        Qubits & \multicolumn{3}{c|}{56} \\
        \hline
        Connectivity & \multicolumn{3}{c|}{All-to-all} \\
        \hline
        Parallel two-qubit operations & \multicolumn{3}{c|}{4} \\
        \hline
        \multicolumn{4}{|l|}{\textbf{Errors}} \\
        \hline
        & \textbf{min} & \textbf{typ} & \textbf{max} \\
        \hline
        Single-qubit gate infidelity & $1 \times 10^{-5}$ & $3 \times 10^{-5}$ & $2 \times 10^{-4}$ \\
        \hline
        Two-qubit gate infidelity & $1.3 \times 10^{-3}$ & $1.5 \times 10^{-3}$ & $3 \times 10^{-3}$ \\
        \hline
    \end{tabular}
    \caption{Quantinuum System H2 Specifications. The data is sourced from the Quantinuum product data sheet \cite{quantinuum_h2_2024}. In our noise model, we use the typical (typ.) calibrated gate infidelity values provided in the data sheet.}
    \label{tab:qtm_parameters}
\end{table}

\begin{table}[htbp]
\centering
\label{tab:gate_error_rates}
\begin{tabular}{@{}cccccc@{}}
\toprule
\textbf{Noise Strength $\lambda$} & 0.5 & 0.75 & 1.0 & 1.25 & 1.5 \\ 
\midrule
\textbf{1q- Gate Error Rate} ($\times 10^{-5}$): $p_1$ & 1.5 & 2.25 & 3 & 3.75 & 6 \\ 
\textbf{2q- Gate Error Rate} ($\times 10^{-3}$): $p_2$ & 0.75 & 1.125 & 1.5 & 1.875 & 2.25 \\ 
\bottomrule
\end{tabular}
\caption{Gate Error Rates at Different Noise Strengths: The single-qubit (1q) and two-qubit (2q) error rates, denoted as $p_1$ and $p_2$, respectively, for noise strength equal to one are derived from the Quantinuum System H2 noise parameters. These error rates are scaled by multiplying with the noise strength factor $\lambda$. The values presented in the table correspond to a composite noise model where the depolarizing noise rate, amplitude damping noise rate, and phase damping noise rate are all equal to the error rates shown.}
\label{tab:ratio_parameters}

\end{table}

\begin{figure}[htbp]
\centering
\begin{subfigure}{0.48\textwidth}
  \centering
  \scalebox{1.0}{
  \Qcircuit @C=1.0em @R=0.2em @!R { \\
        \nghost{ } & \lstick{  } & \gate{\mathrm{H}} & \ctrl{1} & \qw \\
        \nghost{ } & \lstick{  } & \qw & \targ & \qw \\
  \\ }}
  \vspace{0.5cm}
  
  \scalebox{1.0}{
  \Qcircuit @C=1.0em @R=0.2em @!R { \\
        \nghost{ } & \lstick{  } & \gate{\mathrm{H}} & \ctrl{1} & \ctrl{1} & \gate{\mathrm{H}} & \gate{\mathrm{H}} & \ctrl{1} & \qw & \qw\\
        \nghost{ } & \lstick{  } & \qw & \targ & \targ & \qw & \qw & \targ & \qw & \qw\\
  \\ }}
  \caption{Noise scaling by unitary folding}
  \label{fig:zne}
\end{subfigure}
\hfill
\begin{subfigure}{0.48\textwidth}
  \centering  \includegraphics[width=0.9\textwidth]{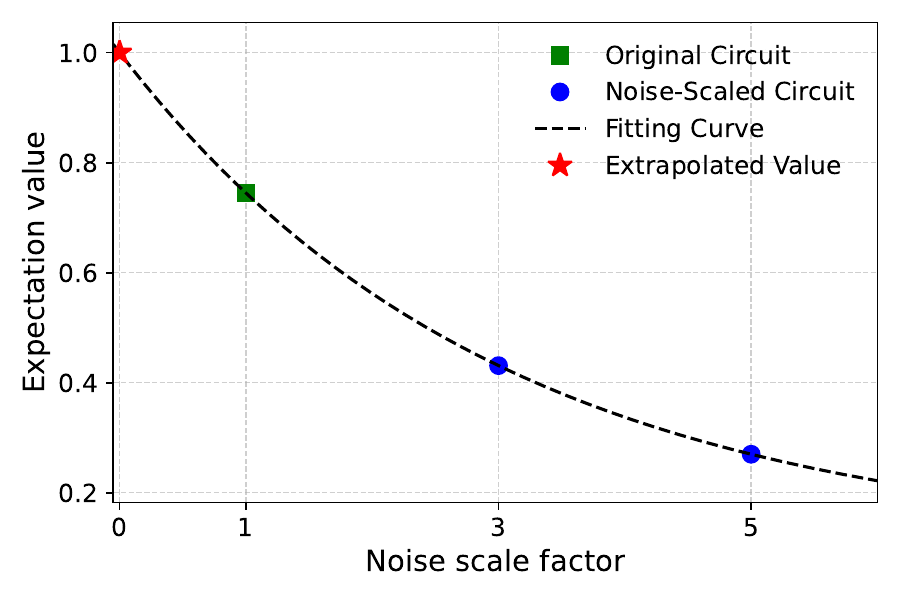}
  \caption{Observable extrapolation}
  \label{fig:extrapolation}
\end{subfigure}
\caption{Zero-noise extrapolation (ZNE) error mitigation technique. (a) Generating a noise-scaled circuit by applying unitary folding. The noise level of the original circuit $U = (\mathrm{H} \otimes I)\,\mathrm{CNOT}$ is intentionally increased by inserting the identity-equivalent operation $U^\dag U$, resulting in $UU^\dag U$. (b) Estimating the zero-noise result by extrapolating noise-scaled results. The unitary folding is performed at different scales (e.g., factors of 3, 5) to obtain multiple noise-scaled expectation values. A parameterized curve (e.g., $a + b e^{-c x}$) is then fitted, and the intercept at zero noise is taken as the error-mitigated result.}
\label{fig: zne_diagram}
\end{figure}

Next, we introduce the working principles and detailed implementation of ZNE in our simulation.

ZNE mitigates noise in quantum circuits by deliberately scaling it through a process known as "circuit folding," where a unitary operation $U$ is transformed into $UU^\dagger U$, as illustrated in Fig. \ref{fig:zne}. This folding operation preserves the ideal unitary evolution while increasing the noise, effectively scaling it by a factor (e.g., approximately three times for a single folding). By varying the folding ratio, we obtain results under different noise regimes. In our simulation, the noise extrapolation factors are set to $1, 1.5, 2, 2.5, 3$. By performing measurements at these different noise levels, we extrapolate the results back to the zero-noise limit using an exponential fit, as shown in Fig. \ref{fig:extrapolation}.

\section{Measurement Sample Complexity}\label{sec:sample_complexity}

In this section, we analyze the sample complexity of the NOQE measurement protocols, both in their original form with the Hadamard test and using shadow tomography. We first establish the necessary preliminaries, including key assumptions on the observable \( O \) and useful trace identities in Sec. \ref{preliminiary}. We then derive the sample complexity of the original NOQE protocol based on Pauli measurements in Sec. \ref{sec: Original NOQE Measurement}, followed by an analysis of improved shadow-based estimation techniques for bilinear functions in Sec. \ref{Shadow Estimation of Specific Nonlinear Functions}. We further analyze the performance using higher order U-statistics for linear and bilinear functions in Sec. \ref{sample_comp_u} and Sec. \ref{appendix:u3_bilinear}. These results highlight the advantages of shadow-based methods over traditional measurement strategies in certain parameter regimes.

\subsection{Preliminaries}\label{preliminiary}

In this section, we consider Hermitian observables \( O \) satisfying the following properties:
\[
\mathrm{Tr}(O^2) \leq B \quad \text{and} \quad \|O\|_\infty = \max_i |\lambda_i(O)| \in \mathcal{O}(1),
\]
where \( \lambda_i(O) \) denotes the eigenvalues of \( O \). That is, the spectral norm is bounded by a constant. This assumption is made without loss of generality, since any observable can be rescaled to satisfy it:
\begin{equation}
    O = \frac{O_{\text{original}}}{\|O_{\text{original}}\|_\infty}.
\end{equation}

As a result of this rescaling, any physical units carried by the observable (e.g., energy units in a Hamiltonian) are absorbed into the scaling factor. The estimation error \( \epsilon \) used throughout this work then refers to the error in estimating the rescaled, unitless observable, and is thus dimensionless. This convention is standard in sample complexity analysis, including in quantum algorithms such as quantum phase estimation. Rescaling in this way helps clearly isolate the dependence on structural parameters like \( B \) and \( \epsilon \) in complexity formulas, without conflating them with physical units. Once a dimensionless estimate is obtained, the corresponding physical quantity (e.g., energy) can be recovered by multiplying the result by \( \|O_{\text{original}}\|_\infty \).

We also note that this implies \( B \leq D \), where \( D \) is the dimension of the Hilbert space on which \( O \) acts.
So one would have $\mathrm{Tr}(O\rho) \leq ||O||_{\infty} \in \mathcal{O}(1)$ and also $\mathrm{Tr}(O^2\rho) \leq ||O||_{\infty}^2 \in \mathcal{O}(1)$. Using the Cauchy-Schwarz inequality, one would also have $\mathrm{Tr}(O)^2 \leq D ~ \mathrm{Tr}(O^2)$.

There are several useful equations that we will use in the derivation
\begin{equation}\label{eq:1}
    \mathrm{Tr}\left((A \otimes B \otimes C \otimes E) (W_{(1,2)} \otimes W_{(3,4)}) (W_{(1,3)} \otimes W_{(2,4)}) \right) =  \mathrm{Tr}(AE)  \mathrm{Tr}(BC),
\end{equation}

\begin{equation}\label{eq:2}
    \mathrm{Tr}\left((A \otimes B \otimes C \otimes E) ~ W_{(3,4)}(W_{(1,3)} \otimes W_{(2,4)}) \right) =  \mathrm{Tr}(AEBC),  
\end{equation}

\begin{equation}\label{eq:3}
    \mathrm{Tr}\left((A \otimes B \otimes C \otimes E) ~ W_{(1,2)}(W_{(1,3)} \otimes W_{(2,4)}) \right) =  \mathrm{Tr}(ACBE),  
\end{equation}

where $W$ denotes the Swap operator.

We also mention that in \ref{appendix:shadow}, \ref{Shadow Estimation of Specific Nonlinear Functions} and \ref{appendix:u3_bilinear}, we only derive the sample complexity for estimating a single function, as the sample complexity for estimating $K$ functions can be obtained from it by using the standard median-of-mean trick.

\subsection{Original NOQE Measurement}\label{sec: Original NOQE Measurement}
The original NOQE protocol uses the generalized Hadamard test to prepare either $|\phi\rangle = |\psi_i\rangle + |\psi_j\rangle$ or $|\phi\rangle = |\psi_i\rangle - |\psi_j\rangle$ depending on the measurement outcome of the control qubit. Then measuring $H$ in the first register gives the expectation value $\langle \phi| H |\phi\rangle = \frac{\langle H\rangle_i + \langle H\rangle_j + 2\mathrm{Re}\langle \phi_i|H|\phi_j\rangle }{2 + 2 \mathrm{Re}\langle\phi_i|\phi_j\rangle}$ thus estimate the $\mathrm{Re}\langle \phi_i|H|\phi_j\rangle$. The imaginary part can be estimated similarly. $\mathrm{Re}\langle\phi_i|\phi_j\rangle$ is given by the probability of getting $+1$ or $-1$ in the control qubit measurement. 

If one wants to estimate $\mathrm{Re}\langle\phi_i|\phi_j\rangle$ to additive error $\epsilon$, one only needs $\frac{1}{\epsilon^2}$ measurements (equivalent to estimate the parameter of Bernoulli distributions), which is typically much smaller than the number of measurements needed to estimate $\langle \phi| H |\phi\rangle$. Thus we can safely ignore the statistical  error in $\mathrm{Re}\langle\phi_i|\phi_j\rangle$.

To estimate $\mathrm{Re}\langle\phi_i|H|\phi_j\rangle$ to an additive error $\epsilon$, we then only need to estimate $\langle \phi |H|\phi \rangle$ to an additive error $\epsilon$ (up to some rescaling because of the presence of $\mathrm{Re}\langle\phi_i|\phi_j\rangle$). The original NOQE protocol measures $H$ by decomposing it into $H = \sum_i w_i P_i$, where $P_i$ is some Pauli operator and $w_i$ is some real coefficient, and then measure individual Pauli terms. Normally, the Frobenius norm of $H$, i.e., the square root of the sum of the absolute squares of all matrix elements is bounded 
\begin{equation}
    \mathrm{Tr}(H^2) = \sum_iw^2_i D \leq B,
\end{equation}
where $D$ is the dimension of the Hilbert space. If $\tilde{N}$ measurements are performed to measure a single Pauli term $P_i$, and the estimator of it is constructed as $\hat{p}_i = n_+ - n_-$ where $n_{\pm}$ is the number of $\pm 1$ in the outcomes, one can check $\mathbb{E} [\hat{p}_i] = \langle\phi|P_i|\phi\rangle$. The variance of it is 
\begin{equation}
    \mathrm{Var}[\hat{p}_i] = \mathrm{E}[\hat{p}_i^2] - (\mathrm{E}[\hat{p}_i])^2 \leq \mathrm{E}[\hat{p}_i^2]  = \frac{1}{\tilde{n}},
\end{equation}
which makes sense because this is again just a Bernoulli distribution estimation. 

After repeating this for every Pauli $P_i$ in $H$, each with $\tilde{n}$ measurements, the estimator $\hat{h}$ for $\langle \phi |H|\phi \rangle$ is then constructed as $\hat{h} = \sum_i w_i \hat{p}_i$. The variance of it is 
\begin{equation}
    \mathrm{Var}[\hat{h}] = \sum_i w_i^2 \mathrm{Var}[\hat{p}]_i \leq \sum_iw_i^2\frac{1}{\tilde{n}} \leq \frac{B}{D\tilde{n}}.
\end{equation}
If one wants to estimate $\langle \phi |H|\phi \rangle$ to additive error $\epsilon$, one needs to ensure $\mathrm{Var}[\hat{h}] \leq \epsilon^2$ (by Chebyshev's inequality). This means $\tilde{n} \geq \frac{B}{D\epsilon^2}$. If there are $\Omega$ Pauli terms in $H$, then the total number of measurements $n$ is 
\begin{equation}\label{NQOE_bound}
    n = \Omega \tilde{n}\geq \mathcal{O}(\frac{B \, \Omega}{D \, \epsilon^2}).
\end{equation}
Similar bounds exist such as $n \geq \mathcal{O}(\frac{(\sum_i |w_i|)^2}{ \epsilon^2})$ \cite{huggins2021efficient}, which is larger than \eqref{NQOE_bound}. Some comments on the bounds in Eq. \eqref{NQOE_bound}, in worst case, $\Omega$ can be as large as $D^2$. In this case $n \geq \mathcal{O}(\frac{B \, D}{ \epsilon^2})$, which is much worse than the bound one would get by performing global Clifford shadow tomography on $|\phi\rangle$, which is $n \geq \mathcal{O}(\frac{B}{ \epsilon^2})$ \cite{grier2024sample, huang2020predicting}.

\subsection{Shadow Estimation of Bilinear Functions}\label{Shadow Estimation of Specific Nonlinear Functions}
Here we derive the number of measurements required to estimate bilinear function of two pure states $\rho$ and $\sigma$ of the form
\begin{equation}
    o = \mathrm{Tr}\left( O \rho \sigma\right) + \mathrm{Tr}\left( O  \sigma \rho\right) = \mathrm{Tr}\left( W\left( I \otimes O + O \otimes I\right) \rho \otimes \sigma \right),
\end{equation}
Where $W$ is the SWAP operator. We perform (global Clifford measurement) shadow tomography on $\rho$ and $\sigma$ respectively, obtaining classical shadows $\{\hat{\rho}_i \}_i^n$ and $\{\hat{\sigma}_i \}_i^n$, each of $n$ samples. We then construct the unbiased estimator for $\rho$ and $\sigma$ as 
\begin{subequations}
\begin{equation}
    \hat{\rho} = \frac{1}{n}\sum_i\hat{\rho}_i
\end{equation}
\begin{equation}
    \hat{\sigma} = \frac{1}{n}\sum_i\hat{\sigma}_i
\end{equation}
\end{subequations}
The unbiased estimator for $o$ is constructed as 
\begin{equation}
    \hat{o} = \mathrm{Tr}\left( W\left( I \otimes O + O \otimes I\right) \hat{\rho} \otimes \hat{\sigma} \right) = \frac{1}{n^2}\sum_{i,k}\mathrm{Tr}\left(W\left( I \otimes O + O \otimes I\right) \hat{\rho}_i \otimes \hat{\sigma}_k \right).
\end{equation}
The accuracy of this estimator is controlled by its variance, which is
\begin{equation}
\begin{aligned}
    \mathrm{Var}[\hat{o}] &= \mathbb{E}[\hat{o}^2] - \mathbb{E}[\hat{o}]^2 \\
    &= \frac{1}{n^4} \sum_{i,j,k,l} \mathrm{Cov}(o_{ik}, o_{jl}) \\
    &= \mathbb{E}\left[\frac{1}{n^4}\sum_{i,j,k,l}\mathrm{Tr}\left(W\left( I \otimes O + O \otimes I\right) \hat{\rho}_i \otimes \hat{\sigma}_k \right)\mathrm{Tr}\left(W\left( I \otimes O + O \otimes I\right) \hat{\rho}_j \otimes \hat{\sigma}_l \right)\right] - o^2 \\
    &= \frac{1}{n^4}\sum_{i,j,k,l}\mathrm{Tr} \left(\left( I \otimes O + O \otimes I\right) \otimes \left( I \otimes O + O \otimes I\right) \mathbb{E}\left[ \hat{\rho}_i \otimes \hat{\sigma}_k \otimes \hat{\rho}_j \otimes \hat{\sigma}_l \right] ~ W_{(1,2)} \otimes W_{(3,4)} \right) - o^2 \\
    &= \frac{1}{n^4}\sum_{i,j,k,l}\mathrm{Tr} \left\{\left( I \otimes I \otimes O \otimes O + I \otimes O \otimes O \otimes I + O \otimes I \otimes I \otimes I + O \otimes O \otimes I \otimes I \right) \right. \\
    &\left. \times \mathbb{E}\left[ \hat{\rho}_i \otimes \hat{\rho}_j \otimes \hat{\sigma}_k \otimes \hat{\sigma}_l \right] ~ W_{(1,3)} \otimes W_{(2,4)} \right\} - o^2
\end{aligned}\label{eq:var_o}
\end{equation}
where the last equality is just relabeling of Hilbert space $2$ and $3$. There is a useful lemma that we will repeatedly use in this note for calculation.
\begin{lemma}{(Lemma 34 of \cite{grier2024sample})}\label{lemma_moment}
    For random Clifford measurement, the second moment of the classical shadow is
    \begin{equation}
        \mathbb{E}[\hat{\rho_i} \otimes \hat{\rho_i}] =(I \otimes I + I \otimes \rho + \rho \otimes I) \left(\frac{D+1}{D+2}W - \frac{1}{D+2} I \right). 
    \end{equation}
\end{lemma}
Now lets evaluate Eq.\eqref{eq:var_o}. The values of various terms in it depend on whether $i$ and $j$, as well as $k$ and $l$ are equal (whether they are independent classical shadows or not). So let's analyze the different situation one by one.

\subsubsection{When $i \neq j$ and $k \neq l$}
In this case $\mathbb{E}\left[ \hat{\rho}_i \otimes \hat{\rho}_j \otimes \hat{\sigma}_k \otimes \hat{\sigma}_l \right] = \rho \otimes \rho \otimes \sigma \otimes \sigma$ and thus $\mathrm{Cov}(o_{ik}, o_{jl}) = 0$

\subsubsection{When $i = j$ and $k \neq l$}
In this case we have 
\begin{equation}
    \mathbb{E}\left[ \hat{\rho}_i \otimes \hat{\rho}_j \otimes \hat{\sigma}_k \otimes \hat{\sigma}_l \right] = \mathbb{E}\left[\rho_i \otimes \rho_i\right] \otimes \sigma \otimes \sigma = (I \otimes I + I \otimes \rho + \rho \otimes I) \otimes \sigma \otimes \sigma \left(\frac{D+1}{D+2}W_{(1,2)} - \frac{1}{D+2} I \right) 
\end{equation}
So we have (now we omit $\otimes$ and indicate explicitly the dot product by $\cdot$ or $\times$ when it exists.)
\begin{equation}
\begin{aligned}
    \mathrm{Cov}(o_{i,k}, o_{j,l}) &= \mathrm{Tr}\left\{(IIOO + IOOI + OIIO + OOII) \cdot (II\sigma \sigma + I \rho \sigma \sigma + \rho I \sigma \sigma) \right. \\
    & \left. \times \left(\frac{D+1}{D+2}W_{(1,2)} - \frac{1}{D+2} I \right) \cdot W_{(1,3)} W_{(2,4)} \right\}  - o^2\\ 
    &= \frac{2(D+1)}{D+2} \left\{\mathrm{Tr}(O\sigma O\sigma) + \mathrm{Tr}(O \sigma  \rho O \sigma) + \mathrm{Tr}(\sigma O^2 \sigma) + \mathrm{Tr}(\sigma O \rho O \sigma) + \mathrm{Tr}(\rho \sigma O^2 \sigma) + \mathrm{Tr}(O \sigma O \rho  \sigma) \right\} \\ 
    &- \frac{4}{D+2}\left\{\mathrm{Tr}(O \sigma)^2 + \mathrm{Tr}(O \sigma)\mathrm{Tr}(\rho O \sigma) + \mathrm{Tr}(O\sigma)\mathrm{Tr}(\sigma O \rho) \right\} - o^2.
\end{aligned}\label{eq:o_ik_o_il}
\end{equation}
Where we used Eq.\eqref{eq:2}. Using the fact that $\rho$ and $\sigma$ are pure states, so that for example $\mathrm{Tr} (O \sigma O \sigma) = \mathrm{Tr}(O \sigma)^2$, we can simply Eq \eqref{eq:o_ik_o_il} further as 
\begin{equation}
\begin{aligned}
     \mathrm{Cov}(o_{i,k}, o_{j,l}) &= \frac{2D-2}{D+2} \left[\mathrm{Tr}(O\sigma)^2 + \mathrm{Tr}(O\sigma) \mathrm{Tr}(O\sigma \rho + O \rho \sigma) \right] + \frac{2D+2}{D+2}\left[|\langle\sigma|O|\rho\rangle|^2 + \mathrm{Tr}(O^2 \sigma)(1 + |\langle \rho | \sigma\rangle|^2) \right] - o^2 \\
    &\leq 2||O||_{\infty}^2 + 4||O||_{\infty}^2 + 2 ||O||_{\infty} ||(I\otimes O + O \otimes I)W||_{\infty} + 4||O||_{\infty}^2 - o^2\\
    &= 10 ||O||_{\infty}^2 + 2||O||_{\infty}\cdot 2||O||_{\infty} -o^2\\
    &= 14 ||O||_{\infty}^2 -o^2\\
    &\leq 14 ||O||_{\infty}^2 \in \mathcal{O}(1)
\end{aligned}
\end{equation}

\subsubsection{When $i\neq j$, $k=l$}
This is similar to the above and we get $\mathrm{Cov}(o_{i,k}, o_{j,l}) \in \mathcal{O}(1)$.

\subsubsection{When $i=j$, and $k = l$}
In this case we have 
\begin{equation}
    \mathbb{E}\left[ \hat{\rho}_i \otimes \hat{\rho}_j \otimes \hat{\sigma}_k \otimes \hat{\sigma}_l \right] = (II + I\rho + \rho I) (II + I\sigma + \sigma I) \times \left(\frac{D+1}{D+2}W_{(1,2)} - \frac{1}{D+2}I \right)\left(\frac{D+1}{D+2}W_{(3,4)} - \frac{1}{D+2}I \right) \times (W_{(1,3)} W_{(2,4)})
\end{equation}
Therefore we have
\begin{equation}
\begin{aligned}
    \mathrm{Cov}(o_{i,k}, o_{l,j}) &= \mathrm{Tr}\left\{(IIOO + IOOI + OIIO + OOII) \right. \\
    &\times (IIII + III\sigma + II\sigma I + I\rho II + I\rho I \sigma + I \rho \sigma I + \rho IIII + \rho II \sigma + \rho I \sigma I) \\
    &\left. \times \left[\frac{(D+1)^2}{(D+2)^2}(W_{(1,2)}W_{(3,4)})  + \frac{1}{(D+2)^2}I - \frac{D+1}{(D+2)^2}\left( W_{(1,3)} + W_{(1,2)} \right) \right](W_{(1,3)} W_{(2,4)})\right\} -o^2\\
    &= \frac{(D+1)^2}{(D+2)^2}\left\{3\mathrm{Tr}(O)^2 + 4 \mathrm{Tr}(O\sigma + O\rho)\mathrm{Tr}(O)  + 4\mathrm{Tr}(O\rho)\mathrm{Tr}(O\sigma) + 2\mathrm{Tr}(O)\mathrm{Tr}(O\sigma \rho + O\rho\sigma)\right. \\
    &\left. + 5\mathrm{Tr}(O^2) + 2D\mathrm{Tr}(O^2\sigma + O^2\rho) + 2\mathrm{Tr}(O^2\rho + O^2\sigma)  + D\mathrm{Tr}(O\rho)\mathrm{Tr}(O\sigma) + 2\mathrm{Tr}(O^2)\mathrm{Tr}(\rho \sigma) + D\mathrm{Tr}(O\rho O\sigma) \right\}\\
    &+ \frac{1}{(D+1)^2}\left\{4 \mathrm{Tr}(O^2) + 8 \mathrm{Tr}(O\rho + O \sigma) \mathrm{Tr}(O) + 4\mathrm{Tr}(O)\mathrm{Tr}(O\rho \sigma + O \sigma \rho) + 8 \mathrm{Tr}(O\sigma)\mathrm{Tr}(O\rho) \right\}\\
    &-\frac{D+1}{(D+2)^2}\left\{ 8 \mathrm{Tr}(O^2) + 16\mathrm{Tr}(O^2\sigma + O^2\rho) + 16 \mathrm{Tr}(O \rho O \sigma) + 8 \mathrm{Tr}(O^2\rho \sigma + O^2 \sigma \rho) \right\} -o^2.
\end{aligned}
\end{equation}
The dominant term of the above expression is the $\mathrm{Tr}(O)^2$ term, which is upper bounded by $D\mathrm{Tr}(O^2)$, therefore, we have
\begin{equation}
     \mathrm{Cov}(o_{i,k}, o_{l,j}) \in \mathcal{O}(D\mathrm{Tr}(O^2)) \leq \mathcal{O}(D B) .
\end{equation}

\subsubsection{Final result}
Putting these all together, we have the bound for the variance of our estimator $\hat{o}$
\begin{equation}
    \mathrm{Var}[\hat{o}] \in \mathcal{O}\left( \frac{n^3 + n^2 dB}{n^4} \right) = \mathcal{O}\left(\frac{1}{n} + \frac{ DB}{n^2} \right).
\end{equation}
Therefore, in order to estimate $o$ within $\epsilon$ with high probability, which means $\mathrm{Var}(\hat{o}) \leq \epsilon^2$, we need the number of samples $n$ to be 
\begin{equation}\label{eq:bound_nonlinear_1}
    n \geq \mathcal{O}\left(\frac{1}{\epsilon^2} + \frac{\sqrt{DB}}{\epsilon} \right).
\end{equation}
Some comments about this bound. Recall that when performing random Clifford measurements on $\rho$, the sample complexity for estimating $\mathrm{Tr}(O\rho)$ is $n\geq\mathcal{O}\left(\frac{B}{\epsilon^2} \right) $, this means that when $\epsilon \leq \sqrt{\frac{B}{D}}$, our bound is smaller than $N\geq\mathcal{O}\left(\frac{B}{\epsilon^2} \right) $ and estimating $\mathrm{Tr}(\sigma O\rho )$ (with $\rho$ and $\sigma$ pure states) is even more accurate than estimating $\mathrm{Tr}(O\rho)$. Another point is that, Eq. \eqref{eq:bound_nonlinear_1} coincides with the sample complexity of estimating $\textit{linear}$ functions of $\rho$ using $m=2$ U-statistics.

\subsection{Shadow Estimation Using U-Statistics}\label{sample_comp_u}
Now we turn to derive the measurement bound for estimating linear functions of the form $\mathrm{Tr}(O\rho)$, where $\rho$ is a pure state, with shadow tomography using U-statistics. First we perform shadow tomography (random Clifford measurement) on $\rho$, obtaining classical shadows $\{ \hat{\rho}_i\}_{i=1}^n$. Now we use order-m U-statistics to construct an unbiased estimator $\hat{\rho}^{(U_m)}$ of $\rho$, which can be taken to be
\begin{equation}\label{eq:u-stat}
    \hat{\rho}^{(U_m)} = \frac{1}{n (n-1)\cdots (n-m+1)}\sum_{i_1 \neq i_2 \neq \dots \neq i_m} \hat{\rho}_{i_1}  \hat{\rho}_{i_2} \cdots  \hat{\rho}_{i_m},
\end{equation}
where $m$ is some positive integer. Since $\rho$ is a pure state ($\rho^m = \rho$), one can check that Eq.\eqref{eq:u-stat} is an unbiased estimator of $\rho$
\begin{equation}
    \mathbb{E}[\hat{\rho}^{(U_m)}] = \rho.
\end{equation}
The original classical shadow \cite{huang2020predicting} which is equivalent to $m=1$ has sample complexity
\begin{equation}
    n \geq \mathcal{O}\left(\frac{B}{\epsilon^2}\right).
\end{equation}
and \cite{grier2024sample} has proved the sample complexity for $m=2$ is 
\begin{equation}
    n \geq \mathcal{O}\left(\frac{1}{\epsilon^2} + \frac{\sqrt{DB}}{\epsilon} \right).
\end{equation}
Now we derive the sample complexity for $m = 3$, which we used in the paper. 

The estimator of our linear function is then
\begin{equation}
    \hat{o} = \mathrm{Tr}(O\hat{\rho}^{(U_m)}).
\end{equation}
The error is controlled by the variance of our estimator, so the rest is devoted to calculating this variance. For $m=3$, the variance is
\begin{equation}
\begin{aligned}
    \mathrm{Var}[\hat{o}] &= \mathbb{E}[\hat{o}^2] - \mathbb{E}[\hat{o}]^2 \\
    &= \frac{1}{[n(n-1)(n-2)]^2} \sum_{i \neq j \neq k} \sum_{\alpha \neq \beta\neq \gamma} \mathbb{E}[\mathrm{Tr}(O \hat{\rho}_i  \hat{\rho}_j  \hat{\rho}_k) \mathrm{Tr}( O \hat{\rho}_\alpha  \hat{\rho}_\beta  \hat{\rho}_\gamma)] - o^2 \\ 
    &=  \frac{1}{[n(n-1)(n-2)]^2} \sum_{i \neq j \neq k} \sum_{\alpha \neq \beta\neq \gamma} \mathrm{Tr}\left\{(O \otimes O ) \mathbb{E}[(\hat{\rho}_i  \hat{\rho}_j  \hat{\rho}_k) \otimes (\hat{\rho}_\alpha  \hat{\rho}_\beta  \hat{\rho}_\gamma)] \right\} - o^2\\
    & = \frac{1}{[n(n-1)(n-2)]^2} \sum_{i \neq j \neq k} \sum_{\alpha \neq \beta\neq \gamma} \mathrm{Cov}(o_{ijk}, o_{\alpha \beta \gamma}).
\end{aligned}
\end{equation}
Let's analyze the terms $\mathrm{Cov}(o_{ijk}, o_{\alpha \beta \gamma})$ case by case.

\subsubsection{When $i, j, k, \alpha, \beta, \gamma$ are all distinct}
In this case, all the terms are independent of the others, and so  $\mathrm{Cov}(o_{ijk}, o_{\alpha \beta \gamma}) = 0$.

\subsubsection{When $i, j, k$ and $\alpha, \beta, \gamma$ has one pair of equal indices}
It is sufficient to consider cases $i = \alpha$, $i = \beta$, $i = \gamma$ or $j = \beta$, as all the other cases are equivalent to one of the above cases.

\subparagraph{$i = \alpha$}
In this case, we need to evaluate
\begin{equation}
    \mathbb{E}[(\hat{\rho}_i  \hat{\rho}_j  \hat{\rho}_k) \otimes (\hat{\rho}_i  \hat{\rho}_\beta  \hat{\rho}_\gamma)] = \mathbb{E}[(\hat{\rho}_i \rho) \otimes  (\hat{\rho}_i \rho)] = \mathbb{E}[\hat{\rho}_i \otimes \hat{\rho}_i ] \times (\rho \otimes \rho).
\end{equation}
Using Lemma \ref{lemma_moment}, we have
\begin{equation}
\begin{aligned}
    \mathrm{Cov}(o_{ijk},o_{\alpha\beta\gamma}) &= \mathrm{Tr}\left\{(O \otimes O) (I \otimes I + I \otimes \rho + \rho \otimes I) \left(\frac{D+1}{D+2}W - \frac{1}{D+2} I \right) (\rho \otimes \rho) \right\}- o^2 \\
    &= \mathrm{Tr}\left\{ \left(\rho O\otimes\rho O + \rho O\otimes\rho O \rho + \rho O \rho \otimes \rho O\right) \left(\frac{D+1}{D+2}W - \frac{1}{D+2} I \right)\right\} - o^2\\
    &\leq 3\mathrm{Tr}(\rho O \rho O) - o^2\\
    & \leq 3\mathrm{Tr}(\rho O)^2\\
    & \leq 3 ||O||_{\infty}^2
\end{aligned}
\end{equation}

\subparagraph{$i = \beta$}
In this case, we need to evaluate
\begin{equation}
    \mathbb{E}[(\hat{\rho}_i  \hat{\rho}_j  \hat{\rho}_k) \otimes ( \hat{\rho}_i  \hat{\rho}_\beta  \hat{\rho}_\gamma)] = \mathbb{E}[(\hat{\rho}_i \rho) \otimes  (\rho\hat{\rho}_i \rho)] = (I \otimes \rho) \times \mathbb{E}[\hat{\rho}_i \otimes \hat{\rho}_i ] \times (\rho \otimes \rho).
\end{equation}
Using Lemma \ref{lemma_moment}, we have
\begin{equation}
\begin{aligned}
    \mathrm{Cov}(o_{ijk},o_{\alpha\beta\gamma}) &= \mathrm{Tr}\left\{(O \otimes O) (I \otimes \rho) (I \otimes I + I \otimes \rho + \rho \otimes I) \left(\frac{D+1}{D+2}W - \frac{1}{D+2} I \right) (\rho \otimes \rho) \right\}- o^2 \\
    & = \mathrm{Tr}\left\{\left( 2\rho O\otimes \rho O \rho + \rho O \rho \otimes \rho O \rho
    \right)\left(\frac{D+1}{D+2}W - \frac{1}{D+2} I \right) \right\}- o^2 \\
    & \leq 3 \mathrm{Tr}(O\rho)^2 - o^2\\
    & \leq 3||O||_{\infty}^2
\end{aligned}
\end{equation}

Similar calculations show that for any case where only one pair of indices are equal, the covariance is always the same order
\begin{equation}
    \mathrm{Cov}(o_{ijk},o_{\alpha\beta\gamma}) \leq \mathcal{O}(||O||_{\infty}^2).
\end{equation}

\subsubsection{When $i, j, k$ and $\alpha, \beta, \gamma$ has two pairs of equal indices }
There are six different situations, which are $(i, j)$ and $(\alpha, \beta)$ are equal; $(i, j)$ and $(\alpha, \gamma)$ are equal; $(i, j)$ and $(\beta, \gamma)$ are equal; $(i, k)$ and $(\alpha, \gamma)$ are equal; $(i, k)$ and $(\beta, \gamma)$ are equal; and $(j, k)$ and $(\beta, \gamma)$ are equal. We need to distinguish two cases in each of the situations. For example, we need to distinguish $i = \alpha$ while $j = \beta$ with $i = \beta$ while $j =  \alpha$, which we call cyclic and anti-cyclic terms respectively. It turns out that the orders of magnitude of the terms only depends on whether they are cyclic or anti-cyclic, but not which one of the six situations they are in. Therefore, here we only show one most convenient example, which is $i = \alpha$ while $j = \beta$ and $i = \beta$ while $j =  \alpha$, to illustrate the calculations.

\subparagraph{$i = \alpha$ while $j = \beta$}
In this case, we need to evaluate
\begin{equation}
    \mathbb{E}[(\hat{\rho}_i  \hat{\rho}_j  \hat{\rho}_k) \otimes ( \hat{\rho}_i  \hat{\rho}_\beta  \hat{\rho}_\gamma)] = \mathbb{E}[(\hat{\rho}_i \hat{\rho}_j \hat{\rho}_k) \otimes  (\hat{\rho}_i \hat{\rho}_j \hat{\rho}_\gamma)] = \mathbb{E}[\hat{\rho}_i  \otimes \hat{\rho}_i]\times \mathbb{E}[ \hat{\rho}_j  \otimes \hat{\rho}_j]  \times (\rho \otimes \rho).
\end{equation}

Using Lemma \ref{lemma_moment}, we have
\begin{equation}
\begin{aligned}
    \mathrm{Cov}(o_{ijk},o_{\alpha\beta\gamma}) &= \mathrm{Tr}\left\{(O \otimes O)  (I \otimes I + I \otimes \rho + \rho \otimes I)^2 \left(\frac{D+1}{D+2}W - \frac{1}{D+2} I \right)^2 (\rho \otimes \rho) \right\}-o^2 \\
    & = \mathrm{Tr}\left\{(\rho O \otimes \rho O) \left(I\otimes I + 3 I \otimes \rho + 3 \rho \otimes I + 2 \rho \otimes\rho \right) \left(\frac{D^2 + 2D + 2}{(D+2)^2}I - \frac{2(D+1)}{(D+2)^2 }W \right)\right\} - o^2\\
    & \leq 9 \mathrm{Tr}(O\rho)^2 - o^2\\
    & \leq 9 ||O||_{\infty}^2
\end{aligned}
\end{equation}

\subparagraph{$i = \beta$ while $j = \alpha$}
In this case, we rearrange the covariance to have 
\begin{equation}
\begin{aligned}
    \mathrm{Cov}(o_{ijk},o_{\alpha\beta\gamma}) &= \mathbb{E}\left[ \mathrm{Tr}\left\{(O \otimes O) (\hat{\rho}_i \otimes  \hat{\rho}_j) ( \hat{\rho}_j \otimes  \hat{\rho}_i) ( \hat{\rho}_k \otimes  \hat{\rho}_\gamma) \right\} \right]\\
    &= \mathbb{E}\left[ \mathrm{Tr}\left\{(O \otimes O) (\hat{\rho}_i \otimes  \hat{\rho}_j) ( \hat{\rho}_j \otimes  \hat{\rho}_i) ( \rho \otimes  \rho) \right\} \right]\\
    &= \mathbb{E}\left[ \mathrm{Tr}\left\{(O \otimes \hat{\rho}_i) (\hat{\rho}_i \otimes  \rho) ( \hat{\rho}_j \otimes  O) ( \rho \otimes  \hat{\rho}_j) \right\} \right] \\ 
    &=  \mathrm{Tr}\left\{(O \otimes I) (\mathbb{E}[\hat{\rho}_i \otimes \hat{\rho}_i]) (I \otimes \rho O ) (\mathbb{E}[\hat{\rho}_j \otimes \hat{\rho}_j]) (\rho \otimes I
    )\right\}.
\end{aligned}
\end{equation}
Using Lemma \ref{lemma_moment}, and ignoring $o^2$ we have
\begin{equation}
\begin{aligned}
    &\mathrm{Cov}(o_{ijk},o_{\alpha\beta\gamma})\\
    &\leq \mathrm{Tr}\left\{(\rho O \otimes I)(I \otimes I + I \otimes \rho + \rho \otimes I) \left(\frac{D+1}{D+2}W - \frac{1}{D+2} I \right) (I \otimes \rho O )(I \otimes I + I \otimes \rho + \rho \otimes I) \left(\frac{D+1}{D+2}W - \frac{1}{D+2} I \right) \right\} \\
    &= \mathrm{Tr}\left\{(\rho O \otimes I + \rho O \otimes \rho + \rho O\rho \otimes I) \left(\frac{D+1}{D+2}W - \frac{1}{D+2} I \right) (I \otimes \rho O + I \otimes \rho O\rho + \rho \otimes \rho O) \left(\frac{D+1}{D+2}W - \frac{1}{D+2} I \right) \right\} 
\end{aligned}
\end{equation}
Let's look at these terms one by one. The term proportional to $\frac{(D+1)^2}{(D+2)^2}$ is 
\begin{equation}
\begin{aligned}
    &\mathrm{Tr}\left\{(\rho O \otimes I + \rho O \otimes \rho + \rho O\rho \otimes I) W (I \otimes \rho O + I \otimes \rho O\rho + \rho \otimes \rho O)W \right\}\\
    &=  \mathrm{Tr}\left\{(I \otimes\rho O + \rho \otimes \rho O  + I \otimes \rho O\rho)  (I \otimes \rho O + I \otimes \rho O\rho + \rho \otimes \rho O) \right\}\\
    &= (4D + 5) \mathrm{Tr}(O\rho)^2.
\end{aligned}
\end{equation}
The term proportional to $\frac{1}{(D+2)^2}$ is 
\begin{equation}
\begin{aligned}
    &\mathrm{Tr}\left\{(\rho O \otimes I + \rho O \otimes \rho + \rho O\rho \otimes I) (I \otimes \rho O + I \otimes \rho O\rho + \rho \otimes \rho O) \right\}\\
    &= 9 \mathrm{Tr}(O\rho)^2.
\end{aligned}
\end{equation}
The term proportional to $\frac{D+1}{(D+2)^2}$ is 
\begin{equation}
\begin{aligned}
    &\mathrm{Tr}\left\{(\rho O \otimes I + \rho O \otimes \rho + \rho O\rho \otimes I) (I \otimes \rho O + I \otimes \rho O\rho + \rho \otimes \rho O) W\right\}\\
    &+ \mathrm{Tr}\left\{(\rho O \otimes I + \rho O \otimes \rho + \rho O\rho \otimes I)W (I \otimes \rho O + I \otimes \rho O\rho + \rho \otimes \rho O) \right\}\\
    &= 18 \mathrm{Tr}(O\rho)^2
\end{aligned}
\end{equation}

Summing these together, we have
\begin{equation}
    \mathrm{Cov}(o_{ijk},o_{\alpha\beta\gamma}) \in \mathcal{O}(D||O||_{\infty}^2)
\end{equation}

\subsubsection{When $i, j, k$ and $\alpha, \beta, \gamma$ has three pairs of equal indices }
In this case, $\alpha, \beta, \gamma$ is just a permutation of $i, j, k$, i.e. $(\alpha, \beta, \gamma) = \pi(i,j,k)$, with $\pi \in S_3$. There are 6 different permutations in $S_3$, and similar to the last subsection, the covariance actually only depends on the parity of $\pi$. Now let's look at the most convenient examples of each of these situations.

\subparagraph{$\mathrm{Sign}(\pi) = +1$ with $i = \alpha$, $j = \beta$ and $k = \gamma$ as the example}
In this case, we have 
\begin{equation}
\begin{aligned}
     \mathrm{Cov}(o_{ijk}, o_{\alpha, \beta, \gamma}) &= \mathrm{Tr}\left\{ (O \otimes O) \mathbb{E}[(\hat{\rho}_i \hat{\rho}_j \hat{\rho}_k) \otimes (\hat{\rho}_i \hat{\rho}_j \hat{\rho}_k)] \right\} - o^2\\
     &= \mathrm{Tr}\left\{ (O \otimes O) \mathbb{E}[\hat{\rho}_i \otimes \hat{\rho}_i] \mathbb{E}[\hat{\rho}_j \otimes \hat{\rho}_j] \mathbb{E}[\hat{\rho}_k \otimes \hat{\rho}_k] \right\} - o^2,
\end{aligned}
\end{equation}
Using Lemma \ref{lemma_moment}, this is just
\begin{equation}
\begin{aligned}
     \mathrm{Cov}(o_{ijk}, o_{\alpha, \beta, \gamma})
     &= \mathrm{Tr}\left\{ (O \otimes O)  (I \otimes I + I \otimes \rho + \rho \otimes I)^3 \left(\frac{D+1}{D+2}W - \frac{1}{D+2} I \right)^3\right\} - o^2 \\
     &= \mathrm{Tr}\left\{(O\otimes O) (I\otimes I + 7I \otimes \rho + 7\rho \otimes I + 12 \rho \otimes \rho)\left(\frac{(D+1)(D^2+2D+4)}{(D+2)^3}W - \frac{3(D+1)^2+1}{(D+2)^3}I \right) \right\} - o^2\\
     &= \frac{(D+1)(D^2+2D+4)}{(D+2)^3}\left\{\mathrm{Tr}(O^2) + 14 \mathrm{Tr}(O^2\rho) + 12 \mathrm{Tr}(O\rho)^2 \right\}\\
     &- \frac{3(D+1)^2+1}{(D+2)^3}\left\{\mathrm{Tr}(O)^2 + 14 \mathrm{Tr}(O)\mathrm{Tr}(O\rho) + 12\mathrm{Tr}(O\rho)^2\right\} - o^2\\
     &\leq \mathrm{Tr}(O^2) + 14 \mathrm{Tr}(O^2\rho) + 12 \mathrm{Tr}(O\rho)^2 + \frac{42}{d+2} |\mathrm{Tr}(O)\mathrm{Tr}(O\rho)|\\
     &\leq \mathrm{Tr}(O^2) + 26 ||O||_{\infty}^2 + 42 \sqrt{\mathrm{Tr}(O^2)/D} ||O||_{\infty}\\
     &\in \mathcal{O}(\mathrm{Tr}(O^2)),
\end{aligned}
\end{equation}

\subparagraph{$\mathrm{Sign}(\pi) = -1$ with $i = \gamma$, $j = \beta$ and $k = \alpha$ as the example}
In this case , we have
\begin{equation}
\begin{aligned}
     \mathrm{Cov}(o_{ijk}, o_{\alpha, \beta, \gamma}) &= \mathrm{Tr}\left\{ (O \otimes O) \mathbb{E}[(\hat{\rho}_i \hat{\rho}_j \hat{\rho}_k) \otimes (\hat{\rho}_k \hat{\rho}_j \hat{\rho}_i)] \right\} - o^2.
\end{aligned}
\end{equation}
Now in order to estimate this, we need the following equation.
\begin{equation}
\begin{aligned}
    (A\otimes C)(B \otimes E) = AB \otimes CE=\operatorname{Tr}_3\left(\left(I \otimes A \otimes E\right)\left(C \otimes B \otimes I\right) W_{(1,3)}\right).
\end{aligned}
\end{equation}
Using this, we have now
\begin{equation}
\begin{aligned}
     \mathrm{Cov}(o_{ijk}, o_{\alpha, \beta, \gamma}) &= \mathbb{E}\left[\mathrm{Tr}\left\{ (\hat{\rho}_k \otimes \hat{\rho}_i)(O\hat{\rho}_i \otimes O\hat{\rho}_k) (\hat{\rho}_j \otimes \hat{\rho}_j) \right\}\right] - o^2\\
     & = \mathbb{E}\left[\mathrm{Tr}\left\{ (I \otimes I \otimes O) (I \otimes \hat{\rho}_k \otimes \hat{\rho}_k) (I \otimes O \otimes I)(\hat{\rho}_i \otimes \hat{\rho}_i \otimes I) W_{(1,3)} (\hat{\rho}_j \otimes \hat{\rho}_j \otimes I) \right\} \right] - o^2\\
     &= \mathrm{Tr}\left\{ (I \otimes I \otimes O) \mathbb{E}[I \otimes \hat{\rho}_k \otimes \hat{\rho}_k] (I \otimes O \otimes I)\mathbb{E}[\hat{\rho}_i \otimes \hat{\rho}_i \otimes I] W_{(1,3)} \mathbb{E}[\hat{\rho}_j \otimes \hat{\rho}_j \otimes I] \right\} - o^2\\.
\end{aligned}
\end{equation}
Now for convenience, we now again omit $\otimes$, and write the dot product explicitly. Using Lemma \ref{lemma_moment} we have
\begin{equation}
\begin{aligned}
     \mathrm{Cov}(o_{ijk}, o_{\alpha, \beta, \gamma})&=
     \mathrm{Tr}\left\{IIO \times (III+II\rho+I\rho I)\times \left(\frac{D+1}{D+2}W_{(2,3)}-\frac{1}{D+2}I \right) \times (IOI) \times (III + I\rho I + \rho I I)\right. \\
     &\left. \times \left(\frac{D+1}{D+2}W_{(1,2)}-\frac{1}{D+2}I \right) \times W_{(1,3)} \times (III + I\rho I + \rho I I) \times \left(\frac{D+1}{D+2}W_{(1,2)}-\frac{1}{D+2}I \right) \right\} -o^2 \\
\end{aligned}
\end{equation}
The dominant term of the above expression is the term proportional to $\frac{(D+1)^3}{(D+2)^3}$, which we evaluate to be
\begin{equation}
\begin{aligned}
    &\frac{(D+1)^3}{(D+2)^3}\mathrm{Tr}\left\{(IIO+II(O\rho)+I\rho O) \times W_{(2,3)} \times (IOI+I(O\rho)I+\rho O I)\times W_{(1,2)}\times W_{(1,3)} \times (III+I\rho I+\rho II) \times W_{(1,2)}\right\}\\
    &=\frac{(D+1)^3}{(D+2)^3}\mathrm{Tr}\left\{(IOI + I (O\rho)I + I O \rho)\times (IOI+I(O\rho)I+\rho O I) \times (III + \rho II + II \rho) \right.\\
    &\left. \times W_{(1,2)} \times W_{(1,3)}\times W_{(1,2)}\times W_{(2,3)} \right\}\\
    &= \frac{(D+1)^3}{(D+2)^3}\mathrm{Tr}\left\{(IOI + I (O\rho)I + I O \rho)\times (IOI+I(O\rho)I+\rho O I) \times (III + \rho II + II \rho) \right\}\\
    & = \frac{(D+1)^3}{(D+2)^3} \left\{(D^2+6D+5)\mathrm{Tr}(O^2) + (2d^2+6d + 5)\mathrm{Tr}(O^2\rho) + (D^2+2D)\mathrm{Tr}(O\rho)^2 + (2D+1)\mathrm{Tr}(O\rho) \right\}\\
    &\in \mathcal{O}(D^2\mathrm{Tr}(O^2))
\end{aligned}
\end{equation}

\subsubsection{Final result}
Putting all together, we have that the variance of the order-$3$ U-statistics estimator is
\begin{equation}
    \mathrm{Var}[\hat{o}] \in \mathcal{O}\left(\frac{n^5+n^4D + n^3 D^2B}{n^6}\right) = \mathcal{O}\left(\frac{1}{n}+\frac{D}{n^2}+\frac{D^2B}{n^3}\right) = \mathcal{O}\left(\frac{1}{n}+\frac{D^2B}{n^3}\right).
\end{equation}
Therefore, to estimate $o$ within error $\epsilon$, we need to make sure the variance is smaller than $\epsilon^2$, which gives the sample complexity
\begin{equation}
    n \geq \mathcal{O}\left(\frac{1}{\epsilon^2} + \frac{D^{2/3} B^{1/3}}{\epsilon^{2/3}} \right).
\end{equation}

\subsection{Shaodw Estimation of Bilinear functions with U-statistics}\label{appendix:u3_bilinear}

Here we discuss the sample complexity of estimating bilinear functions of the form $o = \mathrm{Tr}(O\rho \sigma) + \mathrm{Tr}(O \sigma \rho)$ using order-3 U-statistics, which we used in the main text. The estimator of $o$ is thus 
\begin{equation}
    \hat{o} = \frac{1}{n^2 (n-1)^2 (n-2)^2}\sum_{i\neq j\neq k} \sum_{p\neq q\neq r} \mathrm{Tr}\left\{O \hat{\rho}_i\hat{\rho}_j\hat{\rho}_k \sigma_p \sigma_q \sigma_r\right\} + \mathrm{Tr}\left\{O  \hat{\sigma}_p \hat{\sigma}_q \hat{\sigma}_r \hat{\rho}_i\hat{\rho}_j\hat{\rho}_k \right\}.
\end{equation}

Again, the sample complexity is determined by the variance 
\begin{equation}
\begin{aligned}
    & n^4 (n-1)^4 (n-2)^4 \, \mathrm{Var}(\hat{o})\\
    &= \sum_{i_1\neq j_1\neq k_1} \sum_{i_2\neq j_2\neq k_2} \sum_{p_1\neq q_1\neq r_1} \sum_{p_2\neq q_2\neq r_2} \mathrm{Tr}\left\{\left(I\otimes I \otimes O \otimes O + I \otimes O \otimes O \otimes I + O \otimes I \otimes I \otimes O + O \otimes O \otimes I \otimes I \right) \right.\\
    &\left.\times \mathbb{E}\left [\hat{\rho}_{i_1}\hat{\rho}_{j_1}\hat{\rho}_{k_1} \otimes \hat{\rho}_{i_2}\hat{\rho}_{j_2}\hat{\rho}_{k_2} \otimes \hat{\sigma}_{p_1} \hat{\sigma}_{q_1} \hat{\sigma}_{r_1} \otimes \hat{\sigma}_{p_2} \hat{\sigma}_{q_2} \hat{\sigma}_{r_2} \right] W_{(1,3)} W_{(2,4)}\right\} - o^2
\end{aligned}
\end{equation}
and the derivation of the sample complexity is very similar to \ref{Shadow Estimation of Specific Nonlinear Functions} and \ref{sample_comp_u} but involves much more calculations. Instead of presenting the full lengthy calculation, here we provide a sketch of the derivation and only focus on the main contributing sources of the variance.

The first main contribution to the variance comes from the terms where there only one pair of indices in $i_1, i_2, j_1, j_2, k_1, k_2$ and $p_1, p_2, q_1, q_2, r_1, r_2$ are equal, e.g. $i_1 = i_2$ while $(j_1, j_2, k_1, k_2)$ are distinct and $(p_1, p_2, q_1, q_2, r_1, r_2)$ are also distinct. A single such term contributes $\mathcal{O}(||O||_{\infty}) = \mathcal{O}(1)$ to the variance, and there are in total $\mathcal{O}(n^{11})$ such terms, leading to a total contribution of $\mathcal{O}(n^{11})$ to the variance (ignoring the $\sim 1/n^{12}$ prefactor).

The second main contribution to the variance comes from the terms with $6$ pairs of indices that are equal. We focus on the example where $i_1 = k_2$, $j_1 = j_2$, $k_1 = i_2$ and $p_1 = r_2$, $q_1 = q_2$, $r_1 = p_2$. In this case, the contribution to the  variance (ignoring the $\sim 1/n^{12}$ prefactor) is
\begin{equation}
\begin{aligned}
    &\mathrm{Tr}\left\{\left(I\otimes I \otimes O \otimes O + I \otimes O \otimes O \otimes I + O \otimes I \otimes I \otimes O + O \otimes O \otimes I \otimes I \right) \right.\\
    &\left.\times \mathbb{E}\left [\hat{\rho}_i\hat{\rho}_j\hat{\rho}_k \otimes \hat{\rho}_k\hat{\rho}_j\hat{\rho}_i \otimes \hat{\sigma}_p \hat{\sigma}_q \hat{\sigma}_r \otimes \hat{\sigma}_r \hat{\sigma}_q \hat{\sigma}_p \right] W_{(1,3)} W_{(2,4)}\right\} \\
    &=\mathrm{Tr}\left\{\left(I^{\otimes 6}  \otimes O \otimes O + I^{\otimes 3} \otimes O \otimes I^{\otimes 2} \otimes O \otimes I + I^{\otimes 2} \otimes O \otimes I^{\otimes 4} \otimes O + I^{\otimes 2} \otimes O \otimes O \otimes I^{\otimes 4} \right) \right.\\
    &\left. \times \mathbb{E}\left[(\hat{\rho}_k \times I \otimes I \otimes \hat{\rho}_k) (I \otimes \hat{\rho}_i \otimes \rho_i \otimes I) (I \otimes  I \otimes \hat{\rho}_j \otimes \hat{\rho}_j) \otimes (\hat{\sigma}_r \otimes I \otimes I \otimes \hat{\sigma}_r)(I \otimes \hat{\sigma}_p \otimes \hat{\sigma}_p \otimes I )(I \otimes I \otimes \hat{\sigma}_q \otimes \hat{\sigma}_q) \right] \right.\\
    &\left.\times W_{(1,3)} W_{(2,4)}W_{(5,7)}W_{(6,8)}W_{(3,7)} W_{(4,8)}\right\},
\end{aligned}\label{eq:u3_bilinear_med}
\end{equation}
where in the equality we used the fact
\begin{equation}
    ABC \otimes EFG = \mathrm{Tr}_{1,2} \left\{(C \otimes I  \otimes I  \otimes E)(I  \otimes G  \otimes A  \otimes I) (I  \otimes I  \otimes B  \otimes F)W_{(1,3)} W_{(2,4)} \right\}.
\end{equation}
By using Lemma \ref{lemma_moment}, we are able to identify the leading order contribution of \eqref{eq:u3_bilinear_med} to be 
\begin{equation}
\begin{aligned}
    &\left(\frac{D+1}{D+2}\right)^6 \mathrm{Tr}\left\{\left(I^{\otimes 6}  \otimes O \otimes O + I^{\otimes 3} \otimes O \otimes I^{\otimes 2} \otimes O \otimes I + I^{\otimes 2} \otimes O \otimes I^{\otimes 4} \otimes O + I^{\otimes 2} \otimes O \otimes O \otimes I^{\otimes 4} \right) \right.\\
    &\left. \times (I^{\otimes 4} + I^{\otimes 3} \otimes \rho + \rho \otimes I^{\otimes 3})W_{(1,4)} (I^{\otimes 4}  + I \otimes \rho \otimes I^{\otimes 2}  + I^{\otimes 2} \otimes \rho \otimes I) W_{(2,3)} (I^{\otimes 4} + I^{\otimes 3} \otimes \rho + I^{\otimes 2} \otimes \rho  \otimes I) W_{(3,4)}\right.\\
    &\left. \otimes (I^{\otimes 4} + I^{\otimes 3} \otimes \sigma + \sigma \otimes I^{\otimes 3})W_{(5,8)} (I^{\otimes 4}  + I \otimes \sigma \otimes I^{\otimes 2}  + I^{\otimes 2} \otimes \sigma \otimes I) W_{(6,7)} (I^{\otimes 4} + I^{\otimes 3} \otimes \sigma + I^{\otimes 2} \otimes \sigma  \otimes I) W_{(7,8)}\right.\\
    &\left.\times W_{(1,3)} W_{(2,4)}W_{(5,7)}W_{(6,8)}W_{(3,7)} W_{(4,8)}\right\}\\
    &\leq \mathrm{Tr}\left\{\left(I^{\otimes 6}  \otimes O \otimes O + I^{\otimes 3} \otimes O \otimes I^{\otimes 2} \otimes O \otimes I + I^{\otimes 2} \otimes O \otimes I^{\otimes 4} \otimes O + I^{\otimes 2} \otimes O \otimes O \otimes I^{\otimes 4} \right) \right.\\
    &\left. \times (I^{\otimes 4} + I^{\otimes 3} \otimes \rho + \rho \otimes I^{\otimes 3}) (I^{\otimes 4}  + I \otimes \rho \otimes I^{\otimes 2}  + I^{\otimes 2} \otimes \rho \otimes I)  (I^{\otimes 4} + \rho \otimes  I^{\otimes 3}  + I \otimes \rho  \otimes I^{\otimes 2}) \right.\\
    &\left. \otimes (I^{\otimes 4} + I^{\otimes 3} \otimes \sigma + \sigma \otimes I^{\otimes 3}) (I^{\otimes 4}  + I \otimes \sigma \otimes I^{\otimes 2}  + I^{\otimes 2} \otimes \sigma \otimes I) (I^{\otimes 4} + \sigma \otimes I^{\otimes 3} + I \otimes \sigma  \otimes I^{\otimes 2}) \times W_{(3,8)} W_{(4,7)}\right\}\\
    &\in \mathcal{O}(D^5 \,\mathrm{Tr}(O^2)).
\end{aligned}
\end{equation}
There are $\mathcal{O}(n^6)$ such terms, and thus they contribute $\mathcal{O}(n^6D^5 B)$ to the variance.

It can be shown that all other types of terms contribute the same or a smaller order of magnitude to the variance as $\mathcal{O}(n^{11})$ or $\mathcal{O}(n^6D^5 B)$. Therefore, the total variance is given by
\begin{equation}
    \mathrm{Var}[\hat{o}] \in \mathcal{O}\left(\frac{n^{11} + n^6 D^5 B}{n^{12}}\right) = \mathcal{O}\left(\frac{1}{n} + \frac{D^5 B}{n^6} \right).
\end{equation}
To ensure the estimation within an additive error $\epsilon$, we need $\mathrm{Var}[\hat{o}] \leq \epsilon^2$, which in turn gives the sample complexity
\begin{equation}
    n\in \mathcal{O}\left( \frac{1}{\epsilon^2} + \frac{D^{5/6}B^{1/6}}{\epsilon^{1/3}}\right)
\end{equation}

\clearpage
\bibliography{main}%

\end{document}